\documentclass[11pt,preprint]{aastex}

\newcommand{\kms}{\,km~s$^{-1}$}

\def\spose#1{\hbox to 0pt{#1\hss}}
\def\simlt{\mathrel{\spose{\lower 3pt\hbox{$\mathchar"218$}}
     \raise 2.0pt\hbox{$\mathchar"13C$}}}
\def\simgt{\mathrel{\spose{\lower 3pt\hbox{$\mathchar"218$}}
     \raise 2.0pt\hbox{$\mathchar"13E$}}}

\slugcomment{Accepted to AJ}

\shorttitle{Rotation in Virgo Cluster dE Galaxies}
\shortauthors{Geha et~al.}

\begin{document}


\title{Internal Dynamics, Structure, and Formation of Dwarf Elliptical
Galaxies: II.~Rotating Versus Non-Rotating Dwarfs}

\author{M.\ Geha}
\affil{UCO/Lick Observatory, University of California,
    Santa Cruz, 1156 High Street, Santa Cruz, CA 95064}
\email{mgeha@ucolick.org}

\author{P.\ Guhathakurta\altaffilmark{1,2}}
\affil{Herzberg Institute of Astrophysics, National Research Council
of Canada, 5071 West Saanich Road, Victoria, B.C., Canada V9E 2E7}
\email{raja@ucolick.org}
\altaffiltext{1}{Herzberg Fellow}
\altaffiltext{2}{Permanent address: UCO/Lick Observatory, University
    of California, Santa Cruz, 1156 High Street, Santa Cruz, CA 95064}

\and

\author{R.\ P.\ van der Marel}
\affil{Space Telescope Science Institute, 3700 San Martin Drive, Baltimore, MD 
21218}
\email{marel@stsci.edu}


\begin{abstract}
\renewcommand{\thefootnote}{\fnsymbol{footnote}}
 
We present spatially-resolved internal kinematics and stellar chemical
abundances for a sample of dwarf elliptical (dE) galaxies in the Virgo
Cluster observed with the Keck telescope\footnote{Data presented
herein were obtained at the W.\ M.\ Keck Observatory, which is
operated as a scientific partnership among the California Institute of
Technology, the University of California and the National Aeronautics
and Space Administration.  The Observatory was made possible by the
generous financial support of the W.\ M.\ Keck Foundation.} and
Echelle Spectrograph and Imager.  In combination with previous
measurements, we find that 4 out of 17 dEs have major axis
rotation velocities consistent with rotational flattening, while the
remaining dEs have no detectable major axis rotation.  Despite this
difference in internal kinematics, rotating and non-rotating dEs are
remarkably similar in terms of their position in the Fundamental
Plane, morphological details, stellar populations, and local
environment. We present evidence for (or confirm the presence of)
faint underlying disks and/or weak substructure in a fraction of both
rotating and non-rotating dEs, but a comparable number of
counter-examples exist for both types which show no evidence of such
structure.  Absorption-line strengths were determined based on the
Lick/IDS system (H$\beta$, Mgb, Fe5270, Fe5335) for the central region
of each galaxy.  We find no difference in the line-strength indices,
and hence stellar populations, between rotating and non-rotating dE
galaxies.  The best-fitting mean age and metallicity for our 17 dE
sample are 5~Gyr and $[\rm Fe/H] = -0.3$~dex, respectively, with rms
spreads of 3~Gyr and 0.1~dex.  The majority of dEs are consistent with
solar $[\alpha/\rm Fe]$ abundance ratios.  By contrast, the stellar
populations of classical elliptical galaxies are, on average, older,
more metal rich, and $\alpha$-enhanced relative to our dE sample.  The
line-strengths of our dEs are consistent with the extrapolation of the
line-strength versus velocity dispersion trend seen in classical
elliptical galaxies.  Finally, the local environments of both rotating
and non-rotating dEs appear to be diverse in terms of their proximity
to larger galaxies in real or velocity space within the Virgo Cluster.
Thus, rotating and non-rotating dEs are remarkably similar in terms of
their structure, stellar content, and local environments, presenting a
significant challenge to theoretical models of their formation.

\end{abstract}




\section{Introduction}\label{intro_sec}
\renewcommand{\thefootnote}{\fnsymbol{footnote}}

Dwarf elliptical (dE) galaxies are the dominant galaxy type in nearby
galaxy clusters, accounting for more than 75\% of all objects in these
regions down to a limiting magnitude of $M_V = -14$ \citep*{bin88,
tre02}.  They are characterized by low effective surface brightness
$\mu_{V,\rm eff}>22$~mag~arcsec$^{-2}$ and faint luminosities $M_V \ge
-18$ \citep[for a review see][]{fer94}.  Unlike classical elliptical
galaxies whose surface brightness profiles tend to be well fit by the
de~Vaucouleurs $r^{1/4}$ law \citep{dev48}, dEs have brightness
profiles that are characterized by S\'{e}rsic profiles \citep{ser68}
with indices ranging between $n=1$--3 (where $n=1$ corresponds to an
exponential law and $n=4$ to an $r^{1/4}$ law) making them appear more
diffuse than classical ellipticals of the same total magnitude.  A
direct illustration of this is the discontinuity in surface brightness
between dEs and the low luminosity end of classical elliptical galaxy
sequence (Kormendy 1985; Bender, Burstein, \& Faber 1992; however, see
Graham \& Guzman~2003).

Detailed studies of the internal dynamics of Local Group dEs have been
carried out over the past two decades \citep*[e.g.,][]{ben91}, but,
until recently, only integrated measurements were available for more
distant dEs \citep{pet93}.  The number of dEs outside the Local Group
with spatially-resolved measurements of internal kinematics has
increased dramatically in the last few years \citep*{der01, geh02,
ped02, sim02}.  These observations have revealed an intriguing
diversity of properties: several dEs in the Virgo and Fornax clusters
appear to be roughly consistent with rotational flattening, while
others have no detectable rotation similar to their Local Group
counterparts \citep{ben91}.  This range in rotation velocities is a
newly found feature of cluster dEs that must be explained by
theoretical models of their formation.

Formation scenarios for dE galaxies fall into two broad categories:
(1)~dEs are old, primordial objects, and (2)~dEs have recently evolved
or transformed from a progenitor galaxy population.  Two independent lines of
observational evidence favor the latter category.
Primordial dwarf galaxies are expected to be less strongly clustered than
massive galaxies in the context of hierarchical structure formation
theories, whereas the dE-to-giant galaxy ratio is observed to be higher in
regions of higher galaxy density \citep{fer91}.  The radial velocity
distribution of galaxies in the Virgo Cluster suggests that dEs may be a
recently accreted population \citep*{bot88,con01}.

From simulations, \citet*{moo98} suggest
that galaxy harassment in clusters can morphologically transform a
spiral galaxy into a dE.  Recent observations of
embedded disks in a handful of Virgo and Fornax dEs support this model
\citep*{jer00, bar02, der03}.  Galaxy harassment partially disrupts the
rotational motion of the progenitor galaxy while increasing velocity
dispersions, and is expected to result in
a range of rotational velocities.  It is yet
unclear whether this scenario can reproduce the large fraction of dEs
with extremely low rotational velocities \citep{geh02}.  Another model
suggests that gas-rich dwarf irregular and/or small spiral galaxies
are transformed into dEs through the process of ram pressure stripping
\citep{fab83}.  In this case, dEs would be expected to largely retain
the rotational properties of their progenitors.  The degree of
rotational support present in dE galaxies is thus a strong constraint
on formation models.

We investigate whether or not dE properties such as morphology,
stellar population, and local environment are correlated with the
degree of internal rotation.  A sample of 17 dEs in the Virgo Cluster
forms the basis of this study.  In \S\,\ref{sec_data}, we present Keck
spectroscopy and imaging for eleven dEs and combine these observations
with the six additional dEs previously presented in
\citet{geh02}.  In \S\,\ref{sec_kin}, we discuss the kinematic
profiles of this sample and distinguish between rotating and
non-rotating dEs.  In \S\S\,\ref{sec_fp}--\ref{sec_ids}, respectively,
we examine the position of the rotating and non-rotating sub-samples
in the Fundamental Plane, study galaxy morphologies, and probe stellar
populations using absorption-line-strength indices.  In
\S\,\ref{sec_environ}, we compare the local environments of rotating
and non-rotating dEs within the Virgo Cluster.  Finally, in
\S\,\ref{sec_discuss}, we argue that major axis rotation velocity
appears to be uncorrelated with other internal properties and local
environment for the dEs in our sample, and discuss the broader
implications of these results.

Throughout this paper a Virgo Cluster true distance modulus of $(m -
M)_0 = 30.92$ is adopted, i.e.,~a distance of 15.3~Mpc, as determined
by the {\it Hubble Space Telescope\/}\footnote{Based on observations with the
NASA/ESA Hubble Space Telescope, obtained at the Space Telescope Science
Institute, which is operated by the Association of Universities for Research
in Astronomy, Inc., under NASA contract NAS~5-26555.} ({\it HST\/}) Key
Project on the extragalactic distance scale \citep{fre01}.  Line-of-sight
extinction values are taken from \citet*{sch98} assuming a standard Galactic
reddening law with $R_V = 3.1$.

\section{Data}\label{sec_data}

\subsection{The Sample}

We present Keck spectroscopy and imaging for eleven Virgo Cluster dE
galaxies (in \S\,\ref{sec_spec} and \S\,\ref{sec_im}, respectively)
and combine these with observations of an additional six Virgo Cluster
dEs previously presented in \citet{geh02}.  This sample of 17 Virgo
dEs is then analyzed together throughout the paper.  The general
properties of the full sample are listed in Table~1.  The distribution of the
dEs on the sky and in velocity space are shown in Figure~\ref{fig_vcc}.  The
galaxies lie primarily near the bright end of the dE luminosity
function and were chosen to cover a range of ellipticities.  All but
three of the galaxies discussed in this paper were classified as ``dE'' by
\citet*{bin85}.  These three galaxies are included in our sample as
they lie in the region of the Fundamental Plane typical for dEs
(Figure~\ref{fig_fp}).  Two of these objects, VCC~1386 and VCC~1488,
were classified as normal ellipticals with an uncertainty flag (E:),
but are likely to be misclassified dEs.  The third object, VCC~1036,
was classified as a transition dE/dS0 galaxy; however, as we discuss
in \S\,\ref{sec_morph}, the presence of disks and other substructure
may be fairly common among dEs.  In several of our figures, we include two
additional galaxies, VCC~1122 and VCC~1407 \citep{sim02}.  These are
the only other Virgo Cluster galaxies with published spatially-resolved
internal kinematics outside our sample that are classified as ``dE'' by
\citet{bin85}.

\subsection{Spectroscopy}\label{sec_spec}

Eleven dE galaxies were observed on 2002 March 10--11 using the
Keck~II 10-m telescope and the Echelle Spectrograph and Imager
\citep[ESI;][]{she02}.  The observing setup was identical to that for
the six additional Virgo dEs presented in \citet{geh02}.  Observations
were made in the echellette mode with continuous wavelength coverage
over the range $\rm\lambda\lambda3900$--$11000\mbox{\AA}$ across
10~echelle orders with a spectral dispersion of 11.4\kms~pixel$^{-1}$.
The spectra were obtained through a $0.75'' \times 20''$ slit,
resulting in an instrumental resolution of 23\kms\ (Gaussian sigma)
over the entire spectrum.  The slit was positioned on the major axis
of each galaxy, with the galaxy's center displaced $\sim5''$
from the center of the slit along its $20''$ length.  A summary of
observing parameters is given in Table~2.

The ESI data were reduced using a combination of IRAF echelle and
long-slit spectral reduction tasks.  This is described in detail in
\citet{geh02}; a brief summary is given here.  The sum of a bright
calibration star exposure and a flat field exposure were used to trace
a fiducial point and the ends of the slit for each of the 10~curved
echelle orders.  After overscan and dark frame subtraction, scattered
light was subtracted from individual frames by fitting a smooth
function to the areas outside the apertures corresponding to the
echelle orders (as defined by the tracing of the slit ends).  To
preserve spatial information, the APALL task was used in ``strip''
mode to extract and rectify two-dimensional rectangular strips for
each echelle order.  The rectified orders for both data and
calibration frames were then interpolated to a common spatial pixel
scale of $0.18''$ per pixel.  Data reduction was carried out on these
rectified strips.  Each strip was divided by its corresponding
normalized flat-field image.  Individual exposures were cleaned of
cosmic rays and combined.  Cosmic rays were easily distinguished from
night sky lines even after the slight smearing caused by interpolation
during the rectification process.

A two-dimensional wavelength solution was determined for each
rectified echelle order from a combined Cu/Ar/Hg/Ne arc lamp spectrum.
The wavelength solution was applied to the data (to render vertical
lines of constant wavelength) using either a logarithmic wavelength
interpolation scheme (with 11.4 \kms~pixels) or linear interpolation
(with $0.2 \mbox{\AA}$\,pixels) for the kinematic and line strength
analyses, respectively.  The sky spectrum was determined for each
combined frame from a $\sim 2''$-wide section near the end of the slit
farthest from the galaxy center ($r\sim15''$), and subtracted from the
rest of the two-dimensional spectrum.  The galaxy data were extracted
into one-dimensional spectra in spatial (radial) bins to achieve a
signal-to-noise level of $\rm S/N \ge 10$ per pixel in all radial
bins, while ensuring that the spatial bin size is at least as large as
the FWHM of seeing during the observations ($\sim0.8''$).  The galaxy
continuum flux in each order of each one-dimensional spectrum was then
individually normalized to unity.  Finally, the strips from the
different echelle orders were combined, weighted by the noise frame,
to create a single two-dimensional long-slit spectrum.  The reddest
and two bluest echelle orders were not included in the analysis due to low
signal-to-noise; the final combined spectrum covers
$\rm\lambda\lambda4800-9200\mbox{\AA}$.

The mean line-of-sight velocity and velocity dispersion as a function
of radius were determined using a pixel-fitting method described in
\citet{vdm94}.  As demonstrated in \citet{geh02}, internal velocity
dispersions from this observing setup can be measured down to the
instrumental resolution of 23~km~s$^{-1}$~with an accuracy of
$\sim1$\%.  Velocity profiles were recovered using template stars
ranging in spectral type from G8III to M0III.  The best fitting
template stars, HD48433 and HD40460 (both K1III stars with [Fe/H] =
$-0.13$ and $-0.42$~dex, respectively) were used to recover the
velocity and velocity dispersion profiles presented in
\S\,\ref{sec_kin}.

\subsection{Imaging}\label{sec_im}

Imaging for the above spectroscopic dE sample came from two sources: archival
{\it HST\/} Wide Field Planetary Camera~2 (WFPC2) images were available for
five of the dEs, while the remaining six were imaged with ESI during our
spectroscopic run.  The WFPC2 data, first presented in \citet{mil98} and
\citet{sti01}, consist of $2 \times 230$-s WFPC2 images in the F555W ($V$)
bandpass, except in the case of VCC~1261 where only archival WFPC2 F702W
($R$-band) imaging was available \citep{res01}; photometric parameters for
this galaxy were transformed to
$V$-band assuming $V-R$ = 0.5 \citep{pru98}.  The WFPC2 images were
cleaned of cosmic rays and combined; instrumental magnitudes were
calibrated to standard passbands using the transformations of
\citet{hol95}.  The ground-based ESI images consist of $2 \times 200$-s
$V$-band exposures, with the seeing FWHM ranging between $0.6''-0.7''$.
In imaging mode, ESI has $0.154''$ pixels and a $2'\times 8'$
field-of-view.  The ESI images were bias subtracted and flat
fielded using twilight sky exposures.  Photometric zero-points
were determined based on observations of standard star fields taken on a
different (but also photometric) night; our surface photometry and total
magnitudes agree with previously published values.

Surface brightness profiles were determined for all galaxies using the
IRAF ELLIPSE isophotal fitting routine \citep{jed87}.  Half-light
effective radii ($r_{\rm eff}$) and effective surface brightnesses
($\mu_{\rm eff}$) were calculated by fitting a S\'{e}rsic profile of the
form $I(r) = I_0 \, {\rm exp}[(r/r_0)^{1/n}]$ to the observed $V$-band
data.  A S\'{e}rsic index $n=1$ represents an exponential profile and
$n=4$ is a de~Vaucouleurs law.  The best-fit S\'{e}rsic profile is
determined by non-linear least-squares fitting to the region
$r=1''-20''$; in this region, contributions from the nucleus and
effects of the different spatial resolution of ESI and WFPC2 should be
negligible.  S\'{e}rsic indices, half-light effective radii and effective
surface brightnesses are listed in Table~3.  The average ellipticity
and $B_4$-parameter (discussed in \S\,\ref{sec_morph}) also listed in
this table were determined over the same radial range.

\section{Results}\label{sec_results}

\subsection{Kinematic Profiles}\label{sec_kin}

Kinematic profiles for the eleven Virgo Cluster dEs are presented in
Figures~\ref{fig_rot} and \ref{fig_nrot}.  We have separately plotted
dEs with significant rotation velocities and those which show no
evidence for substantial rotation along the major axis ($v_{\rm rot}
\le 2.5$ \kms, or $v_{\rm rot}/\sigma = 0.1$ see below).  For the
former kinematic profiles (Figure~\ref{fig_rot}), our observations do
not reach out to a large enough radius in three out of four dEs to
observe a turnover in the rotation curve; in these galaxies we measure
only a lower limit to the maximum rotational velocity.  We argue below
that our observations are not far from the turnover radius.  The
maximum rotation velocity ($v_{\rm rot}$) are quoted as lower limits
in Table~1 and shown as up-arrows in Figure~\ref{fig_vsigma}.  For
galaxies which show no evidence for substantial rotation
(Figure~\ref{fig_nrot}), we estimate the upper limit on the maximum
rotation velocity by differencing the average velocities on either
side of the major axis of the galaxy and dividing by two.  Error bars
on rotational motion were determined by adding in quadrature the error
of the mean velocity on either side of the major axis.  These
quantities along with the heliocentric velocity and velocity
dispersion are listed in Table~1.

We note that five dEs in our combined sample (VCC~543, VCC~856,
VCC~1036, VCC~1087, VCC~1261) have been previously observed by
\citet{sim02} and/or \citet{ped02}.  Our kinematic profiles are
entirely consistent with these measurements.  Due to the higher spectral
resolution and signal-to-noise ratio of our observations, we are able
to place significantly tighter constraints on the rotational velocity
for two of these galaxies.  In all cases, we obtain a more accurate
estimate of the velocity dispersions.

For the dE profiles in Figure~\ref{fig_nrot}, it is difficult to
explain the lack of observed major axis rotation as being a result of
insufficient radial coverage.  The majority of our kinematic profiles
are measured out to between 0.5--1.0 $r_{\rm eff}$.  Two-integral
dynamical models predict that if these galaxies were oblate, isotropic
rotators we should have reached or measured beyond the maximum
rotation velocity at these radii \citep[see Figures 8--9 in][]{deh94}.
Although these models are based on the specific case of density
profiles with $\rho \propto r^{-4}$ at large radii and a varying inner
cusp slope, the predictions for $v(r/r_{\rm eff})$ are generic and
should not be very different for a S\'{e}rsic profile (as is more appropriate
for dE galaxies).  As an example, we measure the velocity profile of
the dE3 galaxy VCC~1261 out to its effective radius of $r_{\rm eff} =
7.3''$ (Figure~\ref{fig_nrot}).  As we discuss below, if the observed
flattening in this galaxy was due to rotation, we would expect $v_{\rm
rot}/\sigma = 0.6$, implying a maximum rotation velocity of 30\kms\
given the average dispersion velocity of 45\kms.  This is in stark
contrast to the measured upper limit on rotation in VCC~1261 of
0.5\kms.  

Conversely, for the four profiles with measurable rotation velocities
(Figure~\ref{fig_rot}), it is not surprising that we observe a
turnover in the rotation curve for only one out of four strongly
rotating galaxies (VCC~856).  For these dEs, we only have kinematic
data out to $\sim0.5~r_{\rm eff}$.  The models of \citeauthor{deh94}
suggest that we have barely reached the maximum rotation velocity in
these systems and would need to observe a little further out in radius
in order to see a turnover in these rotation curves.  The rotation
velocities for the these dEs are listed as lower limits in Table~1.

To compare the degree of rotational support in our full sample of 17
dEs, we plot the ratio of the maximum rotational velocity to the
average velocity dispersion ($v_{\rm rot}/\sigma$) versus ellipticity
(Figure~\ref{fig_vsigma}, left panel).  The solid line in this figure
is the ratio expected from an oblate, isotropic,
rotationally-flattened body seen edge-on; systems that are not edge-on
should have somewhat larger predicted $v_{\rm rot}/\sigma$ values
\citep{bin87}.  The lower limits on $v_{\rm rot}/\sigma$ for the four
galaxies with measured rotation are consistent with rotational
flattening.  The upper limits on $v_{\rm rot}/\sigma$ determined for
the majority of our dEs are significantly smaller than expected if the
observed flattenings were due to rotation and imply that these objects
must be supported by anisotropic velocity dispersions.  In our present
sample, the degree of rotational support does not appear to be a
continuous distribution: we observed either dEs with no measurable
major axis rotation, or dEs with rotation velocities approaching that
expected for rotational support.  There is a natural division in our
sample between ``rotating'' and ``non-rotating'' dEs; we arbitrarily
place the dividing line in our sample as galaxies with rotational
support greater or less than $v_{\rm rot}/\sigma = 0.1$.  Rotating and
non-rotating dEs cover a similar range of ellipticities and absolute
magnitudes.  Figure~\ref{fig_vsigma} suggests that rotational support
is not correlated to either of these quantities.  In the remaining
sections, we explore whether or not any other global dE property is
correlated to rotational properties.

The mean velocity dispersions for the dEs in Figures~\ref{fig_rot} and
\ref{fig_nrot}, determined outside of $r = 1''$ to exclude any nuclear
contributions, range between 25 and 45~\kms.  The velocity dispersion
profiles are either constant as a function of radius or show a
decrease in dispersion in the central region.  As demonstrated in
\citet{geh02}, such profiles can in general be adequately modeled
assuming that the stellar mass density profile is equal to the
luminosity density profile times a constant mass-to-light ratio.
Although spherical, isotropic models do not necessarily fit the
observed velocity dispersion profiles in detail, such models do in
general reproduce the overall shape of the observed dE velocity dispersion
profiles.

\subsection{The Fundamental Plane}\label{sec_fp}

Dwarf elliptical galaxies lie in a region of Fundamental Plane space
distinctly different from other stellar systems.  This is best
illustrated in the $\kappa$-projection of the multivariate space
defined by central velocity dispersion, $\sigma_0$, effective surface
brightness, $\mu_{\rm eff}$, and effective radius, $r_{\rm eff}$
\citep{ben92}.  As shown in Figure~\ref{fig_fp}, dEs appear well
separated from other stellar systems.  \citet{ben92} argued that this
separation represents a fundamental difference between the galaxy
formation processes of dEs and more luminous classical ellipticals
\citep[see][however, for an alternate viewpoint]{guz03}.
Here, we are primarily interested in the
relative position of our rotating and non-rotating dE galaxies within
Fundamental Plane space.

In both panels of Figure~\ref{fig_fp}, the rotating and non-rotating
dEs occupy similar regions of the Fundamental Plane.  As a statistical
comparison, we use the one-dimensional Kolmogorov-Smirnov (K-S) test.
This test gives the probability ($P_{\rm KS}$) that the difference
between two distributions would be as large as observed if they had
been drawn from the same population.  Two distributions are considered
different if the probability that they are drawn from the same parent
distribution can be ruled out at a confidence level greater than 95\%
($P_{\rm KS} < 0.05$).  As an example, the probability that the
rotation velocities of our rotating and non-rotating subsamples are drawn
from the same parent distribution is 0.3\% ($P_{\rm KS}$ = 0.003); thus
this hypothesis can be ruled out at high confidence level.  In comparing
$\kappa$-values for rotating and non-rotating dEs (listed in Table~3),
the relatively high values of the K-S probability, $P_{\rm KS}$ = 0.6,
0.2, and 0.1 for $\kappa_1$, $\kappa_2$, and $\kappa_3$ respectively,
indicate that we cannot demonstrate that these two samples are drawn
from different populations.  The rotating dEs appear slightly offset
in the direction of higher $\kappa_1$ and $\kappa_2$ values
(corresponding to larger mass and surface brightness, respectively),
but more data is required to establish such a correlation.

\subsection{Morphology}\label{sec_morph}

Evidence for underlying disk-like structure has been presented for a
handful of cluster dE galaxies based on deep imaging.  Five dEs in the
Virgo Cluster show evidence for a bar or weak spiral structure
\citep{jer00,bar02} and two Fornax Cluster dE/dS0s show evidence for
embedded disks \citep{der03}.  This accounts for roughly 20\% of
analyzed images.  Two of these galaxies are in our sample:
\citeauthor{jer00} discovered faint spiral structure in VCC~856
(IC~3328) implying the presence of a nearly face-on disk, while
\citeauthor{bar02} presented evidence for an elongated dumbbell-shaped
structure along the major axis of VCC~940 (IC~3349).  The former
galaxy is a rotating dE in our sample, the latter is a non-rotating dE.
In view of these results, we analyzed all our dE images to determine
the level of substructure present and whether or not
this is correlated with the amount of major axis rotation.

Images of the dE galaxies in our sample appear to be smooth and
without obvious signs of substructure, however, detailed analysis is
required to detect possible low level substructure.  As discussed in
\S\,\ref{sec_im}, we determine surface brightness profiles using the
IRAF ELLIPSE isophotal fitting routine.  Deviations from the
best-fitting ellipse, often an indication of an underlying disk or
other substructure, are quantified by expanding the intensity
variations along the best-fitting ellipse in a Fourier series as:
\begin{equation}\label{eqn1}
I(\phi) = I_o + \sum_{n=1}^{N} [{\tilde A}_n \sin(n \phi) + {\tilde B}_n \cos(n \phi)].
\end{equation}
\noindent
The first- and second-order terms ($n=1,2$) of Eqn.~(\ref{eqn1}) are zero
by definition; non-zero values would indicate that the best-fitting
ellipse has not been found.  For $n\ge3$, the quantities
\begin{eqnarray}
  A_n = {\tilde A}_n / [a (dI/da)], \\
  B_n = {\tilde B}_n / [a (dI/da)],
\end{eqnarray}
\noindent
measure deviations of the isophote from the best-fitting ellipse,
where $a$ is the semi-major axis length and $(dI/da)$ the local
intensity gradient. The quantity $B_4$ is similar to the quantity
$a_4/a$ used by e.g., \citet*{ben88}. Positive $B_4$ values indicate
disky isophotes whereas negative values indicate boxy isophotes
\citep{jed87}.  In Figure~\ref{fig_sbesi}, we present surface
brightness, ellipticity, position angle, and $B_4$ as a function of
radius for those dEs with ESI~imaging.  We note that in all cases the
isophotal ellipticity varies with radius; two galaxies show
significant departures from zero in their $B_4$ profiles.  For the
remaining dEs in our sample for which we have only WFPC2 imaging, we
present the averaged ellipticity and $B_4$ measurements in Table~3.
We do not present the WFPC2 radial profiles since much of this data
has been presented elsewhere \citep{geh02, sti01, geh03}, and because
these images are too shallow to meaningfully constrain the amount of
substructure present.  As an illustration, the existing WFPC2 images
do not show detectable substructure in VCC~940; however, as mentioned
above, \citet{bar02} present clear evidence for an elongated structure
along its major axis from deep ground-based imaging.

In the first column of Figure~\ref{fig_imesi} we present $V$-band ESI
images for the profiles presented in Figure~\ref{fig_sbesi}.  In the
second column of this figure we plot the residual image resulting from
subtraction of a 2D model based on the ellipse fitting (excluding
higher-order components) from the original image.  In the third column
we present unsharp-masked images created by subtracting a
boxcar-smoothed image (with $4''$ smoothing length) from the original,
to highlight high frequency spatial structure.  The unsharp-masked
images have the advantage of being independent of our ellipse fitting
model, but can be difficult to interpret.  For example, the
``structure'' apparent in the unsharp-masked images of VCC~1947 and
VCC~745 is due to ellipticity gradients evident in
Figure~\ref{fig_sbesi}; the model-subtracted images of these galaxies in the
middle column of Figure~\ref{fig_imesi} show no obvious structure.  On
the other-hand, the model-subtracted images of two other dEs in
Figure~\ref{fig_imesi}, VCC~1036 and VCC~1488, do show clear signs of
an underlying disk.  This is seen in both the unsharp-masked images
and the positive $B_4$ profiles in Figure~\ref{fig_sbesi}.  The
rotating dE/dS0 VCC~1036 has a strong disk component, while the
non-rotating dE VCC~1488 has a weaker underlying disk.  Both these
disks lie along the major axis defined by the outer isophotes.  In
addition, we confirm the faint spiral structure in VCC~856 first
presented by \citet{jer00}; this is strong evidence for a nearly
face-disk.  The remaining three dEs do not show any residuals.  Most
notably, the strongly rotating dE VCC~1947 shows no evidence for a
disk component.

In summary, there is evidence for underlying disks and/or substructure
in both rotating and non-rotating dEs, but counter examples also exist
for both kinematic types which show no evidence of such structure.
For the rotating dEs, we present or confirm evidence for underlying
disk structure in two dEs, but observe no evidence for any
substructure in a third rotating dE.  For the non-rotating dEs, we
present evidence for a weak disk in one galaxy, in addition to one
non-rotating dE with substructure presented in the literature.  Two
non-rotating dEs show no evidence for underlying structure.  For the
remaining dEs in our sample, WFPC2 imaging is not deep enough to place
limits on the presence of substructure.  We conclude that there can be
interesting substructure in dEs, but its presence or absence does not
appear to be a good indicator of the observed dynamics.

\subsection{Line-Strength Indices}\label{sec_ids}

We compute line-strength indices according to the Lick/IDS system
\citep{wor94} in the wavelength region $4800 - 6000\mbox{\AA}$.  This
index system is calibrated to a fixed spectral resolution of $\sim
8\mbox{\AA}$.  To match indices we must degrade the spectral
resolution of our spectra.  Although this means sacrificing a wealth
of fine structure almost certainly sensitive to stellar populations,
higher spectral resolution stellar population models do not yet exist
which cover a sufficient range of ages, metallicities and wavelengths.
The adopted approach also allows us to compare directly to the measured indices
of more luminous elliptical galaxies.

Line-strength indices were computed by first shifting the spectra to
rest frame wavelengths and convolving with a wavelength-dependent
Gaussian kernel to match the spectral resolution of the Lick/IDS
system.  Line-strengths were then calculated according to the index
definitions of \citet{wor94}.  The size of the smoothing kernel was
determined by matching the measured indices of 16 stars (spectral type
F9IV--M0III) observed through the Keck/ESI setup to the values of
\citet{wor94}.  The size of the kernel ranged between $8-9\mbox{\AA}$
(40--45 ESI spectral pixels, Gaussian sigma); the average residual
between measured and published indices was $0.15\mbox{\AA}$.  The
measured line-strengths for these standard stars are plotted against
the published values in Figure~\ref{fig_idscal}.  Small zero-point
corrections due to systematic effects such as sky subtraction and flat
fielding errors both in our data and the published systems
\citep{wor97} are applied to the measured indices of each galaxy.
These corrections were determined from the best fitting offset between
our measured standard star indices and the published values assuming a
linear relation with unit slope.  These corrections are listed in
each panel of Figure~\ref{fig_idscal} and range between 0.1 --
$0.2\mbox{\AA}$.  We do not correct for line broadening due to the
intrinsic velocity dispersion of each dE galaxy as these are much
smaller than the Lick/IDS broadening function.

Error bars on the indices were computed via Monte-Carlo simulation.
We added noise to a high signal-to-noise template star based on the
noise spectrum of each individual galaxy and compute indices with the
same procedure described above.  Error bars were determined for each
index of each galaxy from the rms value after 1000 noise realizations.
Error bars include contributions from photon noise, read-out noise and
pixel-to-pixel correlations due to Gaussian smoothing.

We present line-strength indices for our Virgo Cluster dE galaxies
measured from spectra extracted in a $0.75'' \times 3''$ aperture
centered on each object.  These spectra have signal-to-noise ratios
ranging between 50--100 per $\mbox{\AA}$.  In Table 4, we list the
central line-strength indices and error bars for H$\beta$, Mgb,
Fe5270, and Fe5335 (see \citeauthor{wor94} for index definitions).  In
addition, we list the combined iron index $\langle\rm Fe\rangle$
defined as (Fe5270 + Fe5335)/2.  Our results for the central regions
of VCC~1073 and VCC~543 agree, within the errors, with previously
published values of H$\beta$ and $\langle \rm Fe\rangle$
\citep{ped02}.  We restrict our line-strength analysis to the central
$r<1.5''$ to avoid galaxy light contamination in our sky spectrum.
Based on the surface brightness profiles, we estimate the level of
this contamination to be a few percent or less in the galaxy center,
increasing to $\sim25\%$ in the outer regions.  A few percent
contamination should not affect the central line-strength values or
the kinematic profiles, but could lead to significant errors in the
line strengths at large radii where the contamination level is
larger.

In Figures~\ref{fig_alpha} and \ref{fig_ids}, we present line-strength
index diagrams comparing the H$\beta$, Mgb and $\langle\rm Fe\rangle$
indices for our dE sample.  The majority of dEs are confined to a
small region in all three panels with some amount of intrinsic scatter
between galaxies.  The exception is VCC~1488, a non-rotating galaxy
with noted non-axisymmetric features (\S\,\ref{sec_morph}), which has
a significantly higher H$\beta$ measurement.  The line-strength
indices of the rotating and non-rotating dEs are similarly distributed
in each index-index diagrams.  Based on the K-S test discussed in
\S\,\ref{sec_fp}, we compute the probability that, for each index, the
line-strengths of these two populations are drawn from different
distributions.  The K-S probability is greater than $P_{\rm KS} = 0.5$
for all our measured indices, meaning the line-strengths of these two
samples are statistically consistent with being drawn from a single
distribution.  We conclude that rotating and non-rotating dEs have
similar stellar populations despite very different internal
kinematics.

We next compare our dE line-strength indices to the single-burst
stellar population models of \citet*{tho02}.  These models predict
line-strength indices for a wide range of metallicities ($-2.25 \le
[\rm Fe/H] \le +0.67$~dex) and include predictions for both solar and
non-solar element abundance ratios.  In Figure~\ref{fig_alpha}, we
investigate the [$\alpha/\rm Fe$] ratio of our dE sample by comparing
the measured Mgb and $\langle \rm Fe \rangle$ indices to model
predictions.  In this diagram, effects due to age and metallicity are
largely degenerate and sensitivity to abundance ratio is maximized. We
assume that Mg traces the $\alpha$-elements.  The majority of dEs in
Figure~\ref{fig_alpha} are consistent with solar abundance ratios,
with some scatter towards both the sub- and super-solar abundance
ratios between $-0.3 \le [\alpha/\rm Fe] < +0.3$~dex which cannot be
explained by differences in age and metallicity alone.  We again note
that the rotating and non-rotating dEs have similar inferred abundance
ratios.  \citet{gor97} first suggested that Virgo dEs are consistent
with solar abundance ratios based on a smaller dE sample.  This is in
marked contrast to the sample of classical elliptical galaxies, taken
from \citet{tra00}, also plotted in Figure~\ref{fig_alpha} which have
super-solar abundance ratios $[\alpha/\rm Fe] \sim +0.3$.  This
difference in abundance ratios can be interpreted as a difference in
the time-scale of star formation: $\alpha$-elements are created
rapidly by Type II supernovae while iron is produced by supernovae
Type Ia on longer timescales.  It can then be argued that the bulk of
star formation in classical ellipticals occurred on much shorter
timescales as compared to dwarf ellipticals.  Our observations are
further evidence that dwarf and classical elliptical galaxies have
very different star formation histories.

To determine the luminosity-weighted stellar ages and metallicities
for our dE sample, we plot the Mgb and $\langle \rm Fe \rangle$
indices against H$\beta$ in Figure~\ref{fig_ids}.  Note that while
conclusions based on relative line-strength differences between dE
galaxies are robust, the derived absolute ages and metallicities may
be affected by unknown systematic errors.  The best fitting age and
metallicity was determined for each galaxy by simultaneously
minimizing the residuals between the observed line-strengths and the
predicted Mgb, $\langle \rm Fe \rangle$, and H$\beta$ indices from the
\cite{tho02} solar abundance ratio models.  The best fitting age and
metallicity for our 17 dE sample are 5~Gyr and $[\rm Fe/H] =
-0.3$~dex, with rms spreads of 3~Gyr and 0.1~dex, respectively. This
is consistent with the ages and metallicities implied by the
dynamically determined $V$-band mass-to-light ratios,
$3~\le~\Upsilon_V~\le~6$, calculated in \citet{geh02} for six dEs in
the present sample; the single-burst population models of
\citet{wor94b} predict $V$-band mass-to-light ratios in this range for
the ages and metallities stated above.  Compared to the
\citeauthor{tra00} sample of classical elliptical galaxies plotted in
Figure~\ref{fig_ids}, the average dE in our sample has stronger
H$\beta$ indices and weaker Mgb and $\langle \rm Fe \rangle$ indices,
implying these dEs are younger and more metal poor than a typical
classical elliptical galaxy.

Despite the clear separation in the Fundamental Plane (see
\S\,\ref{sec_fp}) between dEs and classical elliptical galaxies,
\citet*{ben93} first noted that the magnesium line-strengths of these
two galaxy types follow a tight trend with velocity dispersion.
\citet*{cal03} investigated this correlation for a sample of
early-type galaxies which included a much larger number of galaxies
with velocity dispersions less than $\sigma = 100$\kms.  They confirm
this correlation for Mgb and several other Lick/IDS indices.  In
addition, the scatter in the line-strengths of H$\beta$, Mgb, and
Fe5270 was found to be in excess of the measurement error for
velocity dispersions less than $\sigma = 100$\kms.  In
Figure~\ref{fig_mgsig}, we plot line-strengths as a function of
velocity dispersion.  We fit a linear relation to the classical
elliptical sample of \citet{tra00} and compare the result to the
positions of the dEs in our sample.  The linear fits to classical
ellipticals, extrapolated to lower velocity dispersions, are
consistent with our measured dE line-strengths.  Although the observed
scatter for our dE sample is larger than that of the classical
ellipticals, our sample size is too small to make a quantitative
statement regarding the intrinsic scatter.

\section{The Local Environment of Rotating versus Non-Rotating 
dEs}\label{sec_environ}

\begin{figure}
\plottwo{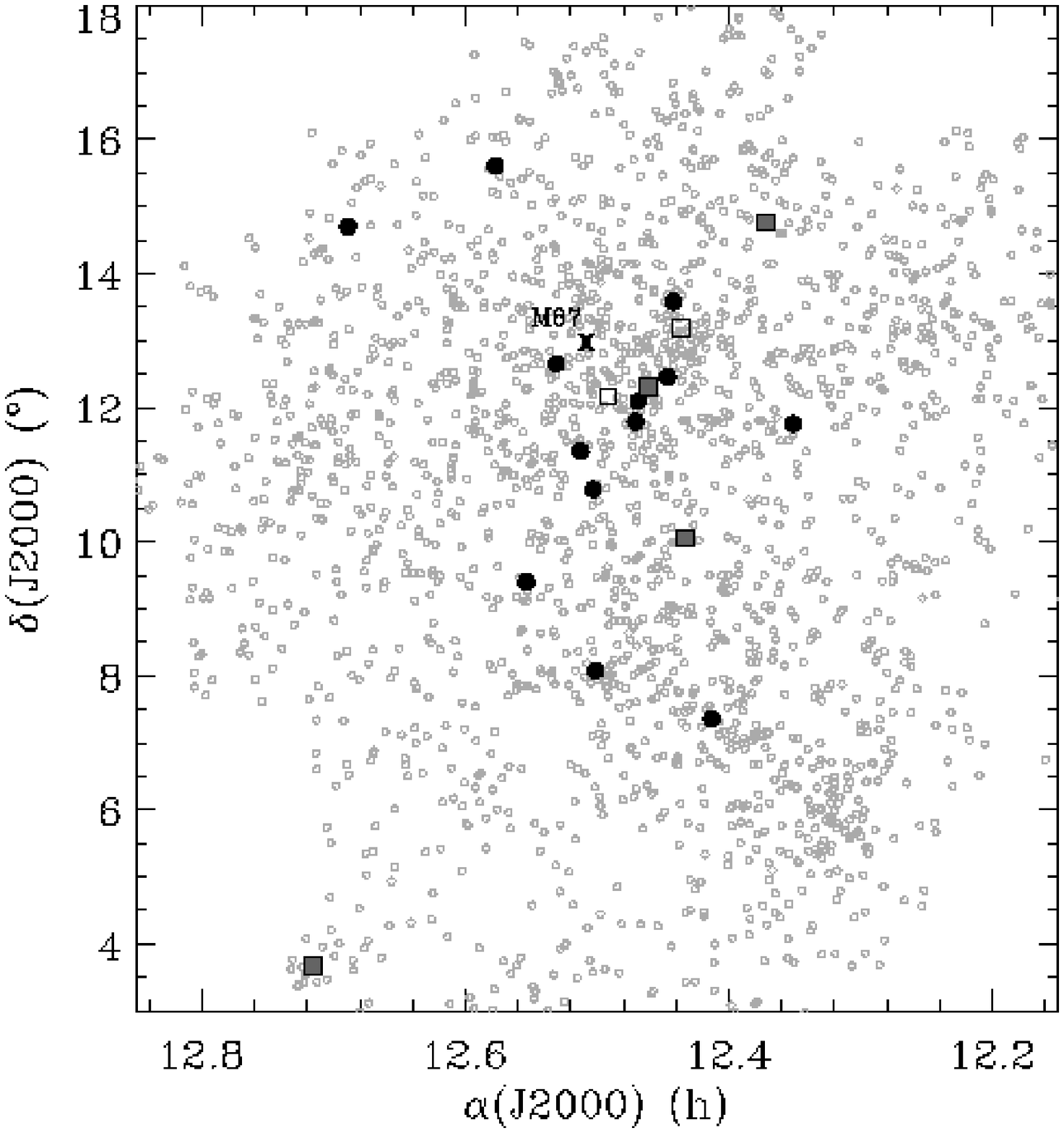}{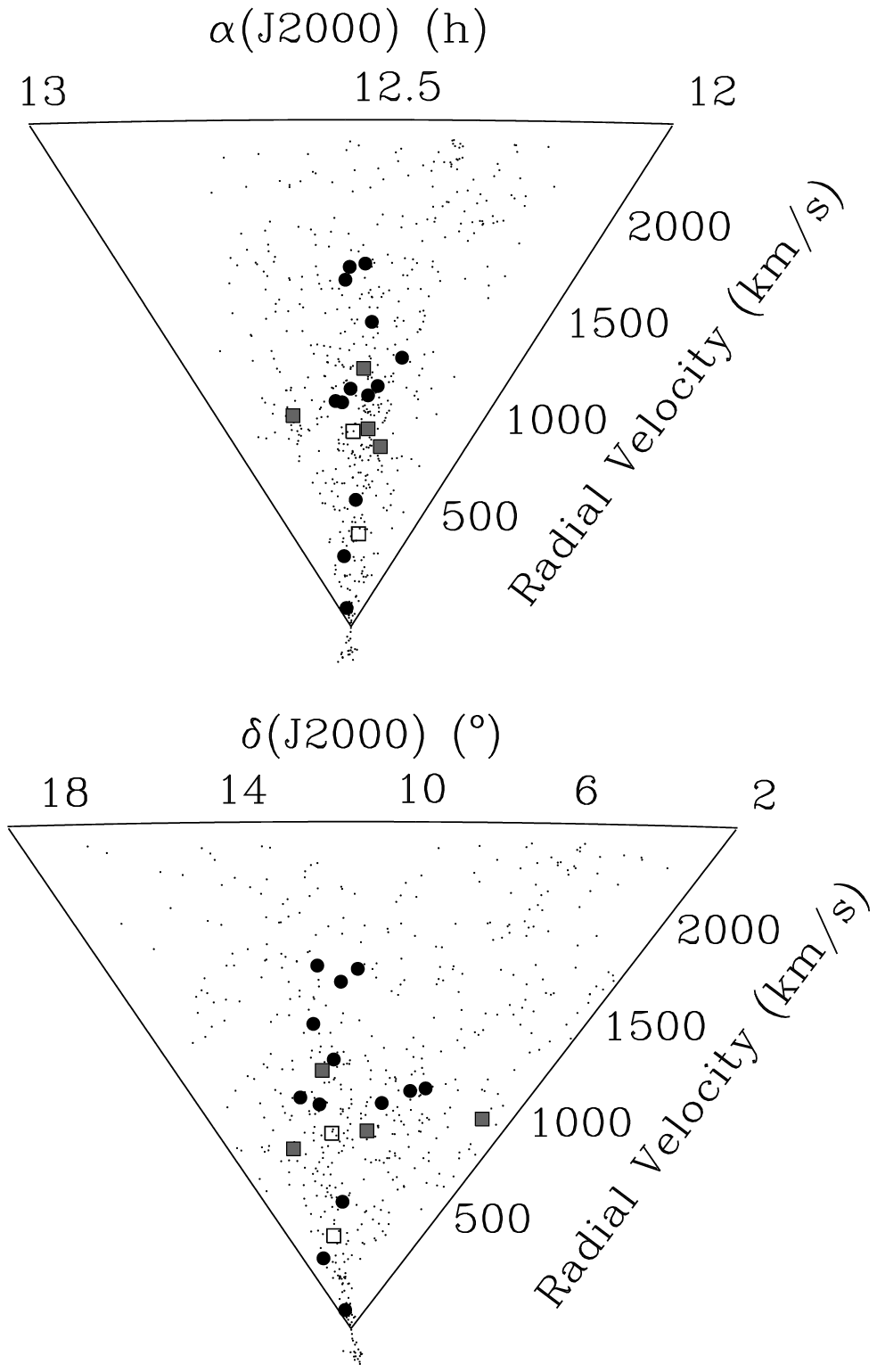}
\vskip 1 cm
\caption{({\it Left panel\/}) Positions (J2000) of the target Virgo
Cluster dE galaxies on the sky relative to all cluster members (grey
open circles) as identified by \citet{bin85}.  Circles
indicate dEs with no detectable rotation, while squares are dEs flattened by
rotation; solid symbols indicate galaxies we have observed with
Keck/ESI, while open symbols are the two additional Virgo dEs, VCC~1122 and
VCC~1407, with kinematic profiles from \citet{sim02}.  The ``X'' symbol
indicates the position of M87 at the center of the Virgo cluster.
({\it Right panels\/}) The same galaxies in velocity-vs-position ``pie''
diagrams.  Small dots represent all cluster members.  \label{fig_vcc}}
\end{figure}

\begin{figure}
\epsscale{0.85}
\plottwo{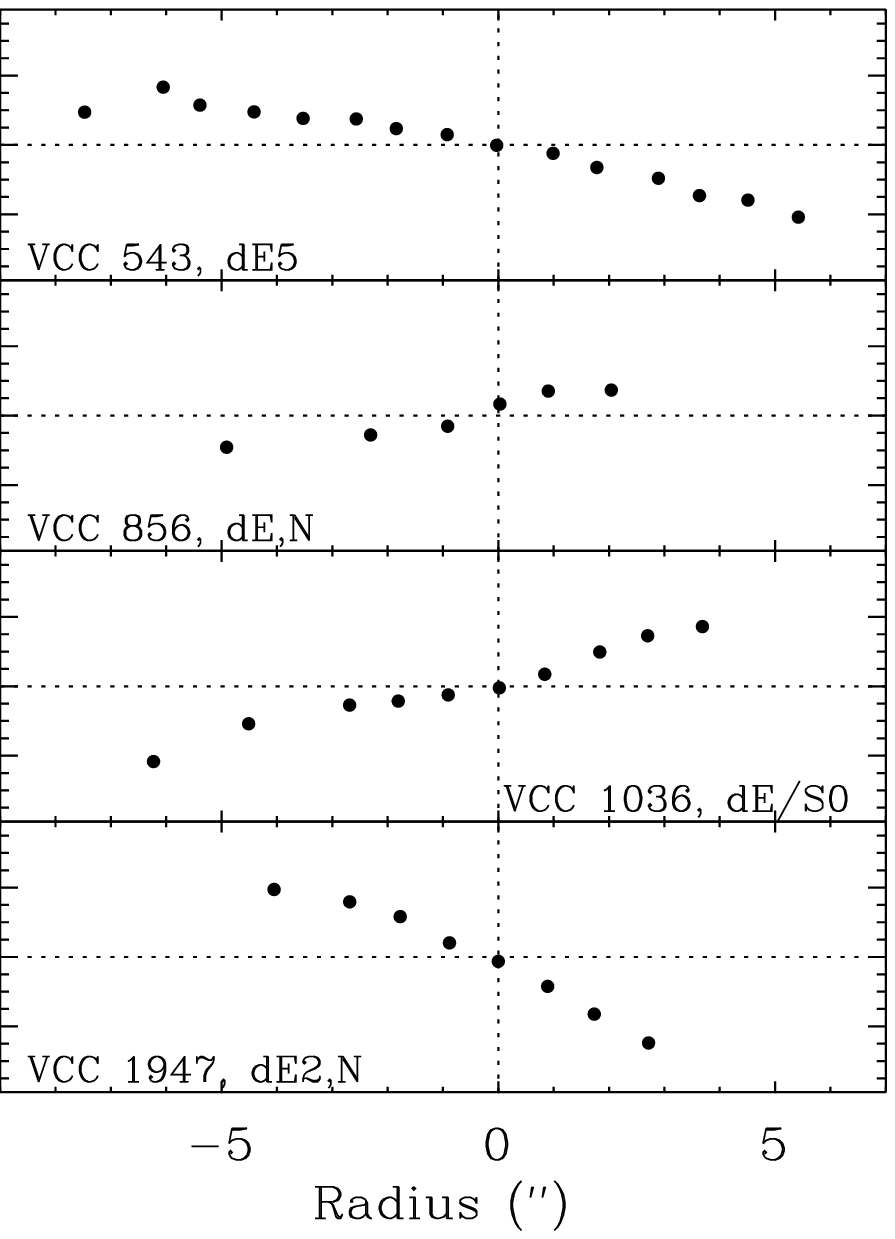}{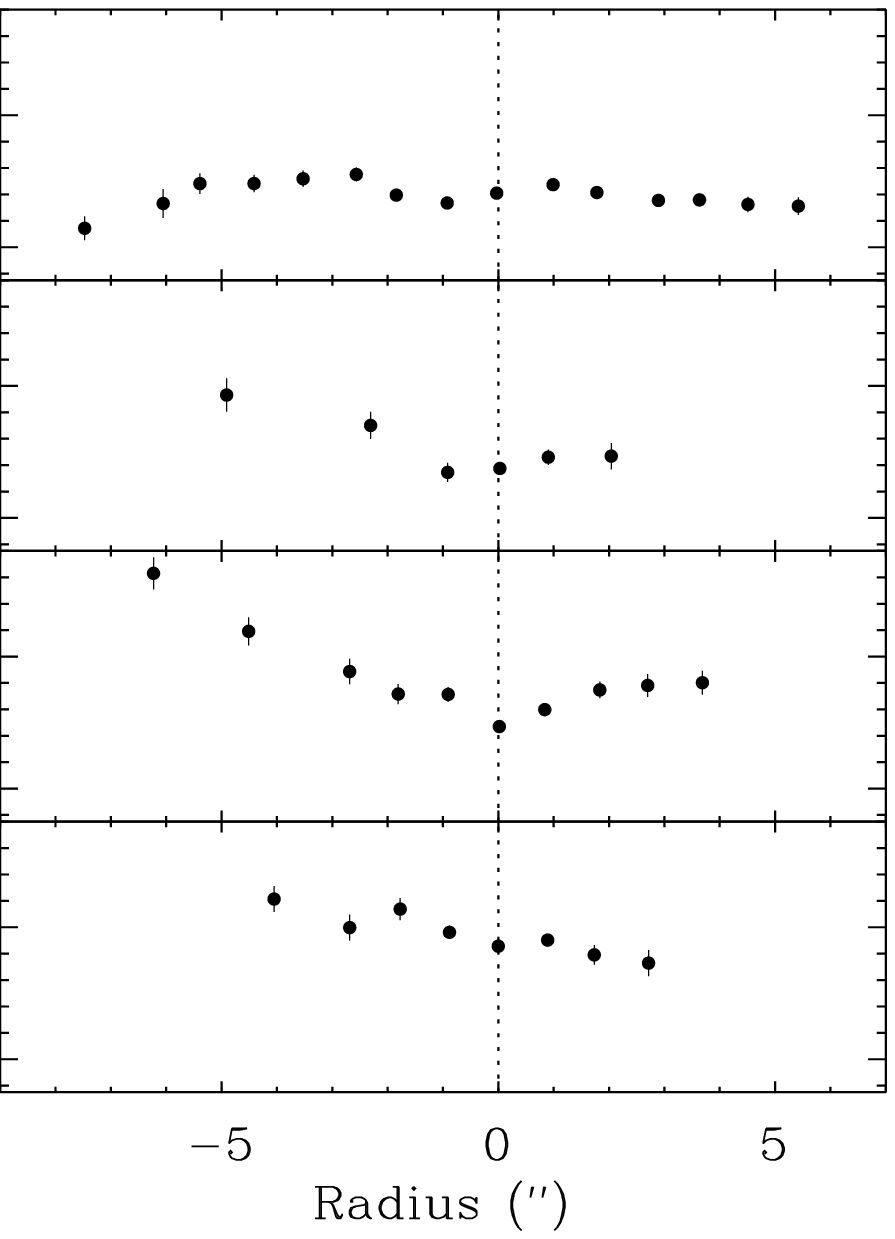}
\vskip -4 cm
\caption{Kinematic profiles for Virgo Cluster dEs with significant
major axis rotation. The mean line-of-sight velocity offset relative
to the galaxy's systemic velocity ({\it left panel\/}) and velocity
dispersion ({\it right panel\/}) are plotted as a function of radial
distance along the major axis.  At the distance of Virgo, $1''$
corresponds to $\sim100$~pc.  One sigma error bars are plotted in both
panels, but they are often smaller than the plotted data points.\label{fig_rot}}
\end{figure}

\begin{figure}
\epsscale{0.85}
\plottwo{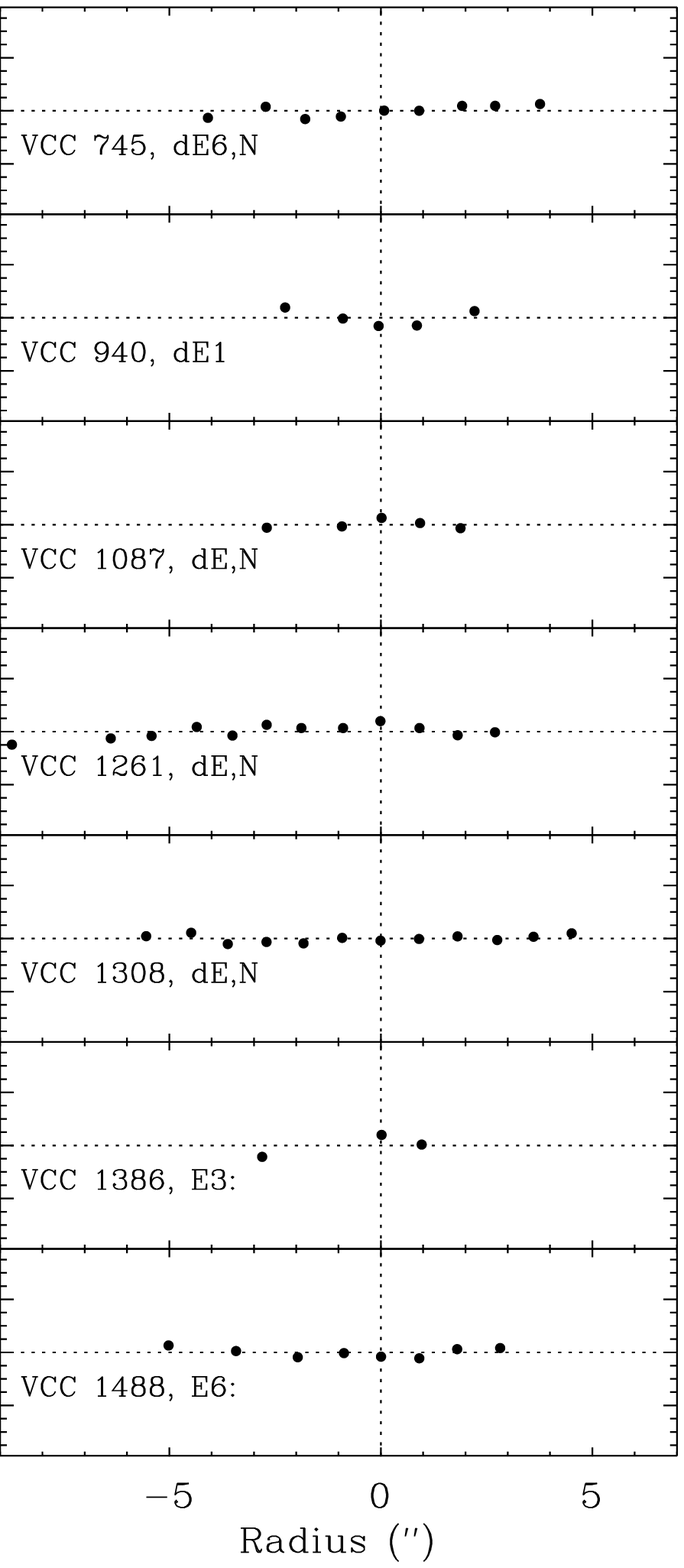}{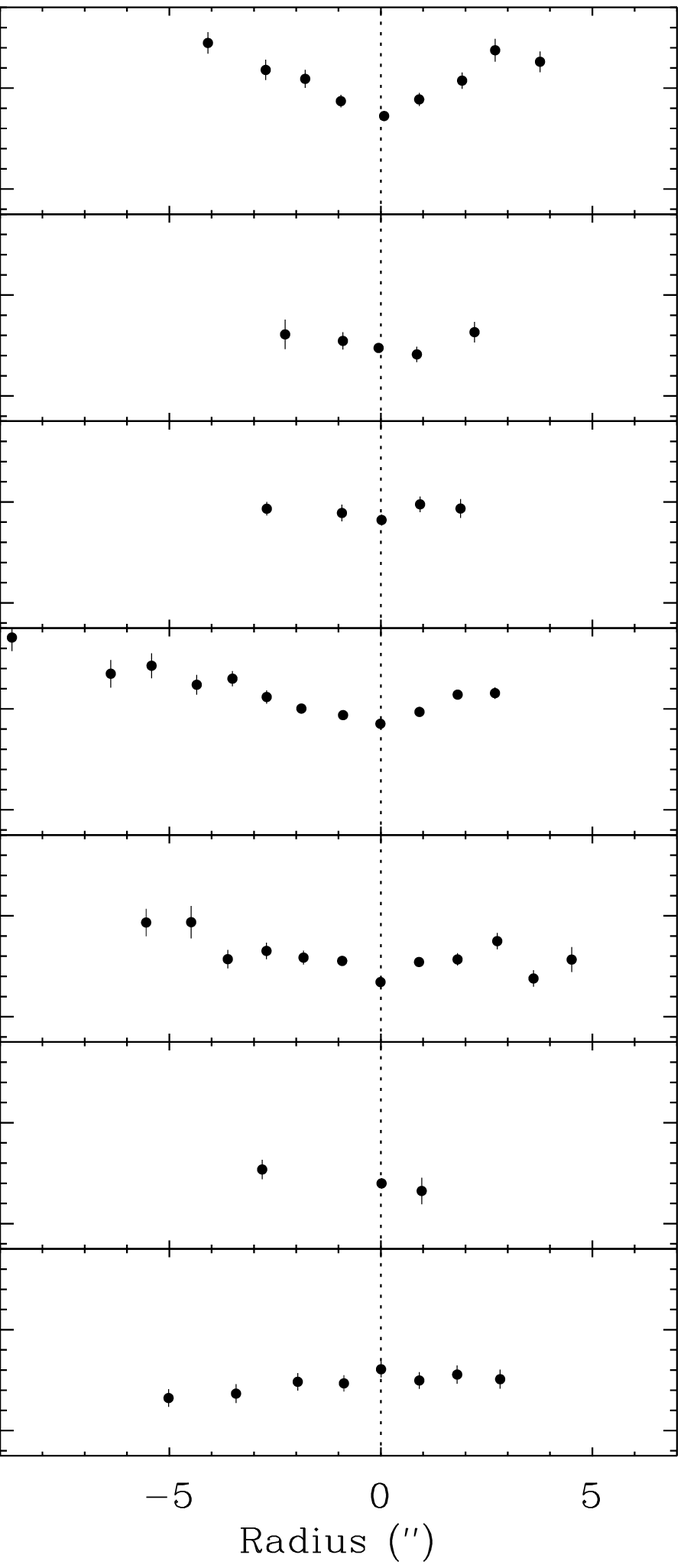}
\vskip 0.5 cm
\caption{Same as Figure~\ref{fig_rot} for those dEs which show no
evidence for substantial rotation along the major axis.\label{fig_nrot}}
\end{figure}

\begin{figure}
\epsscale{1.0}
\plotone{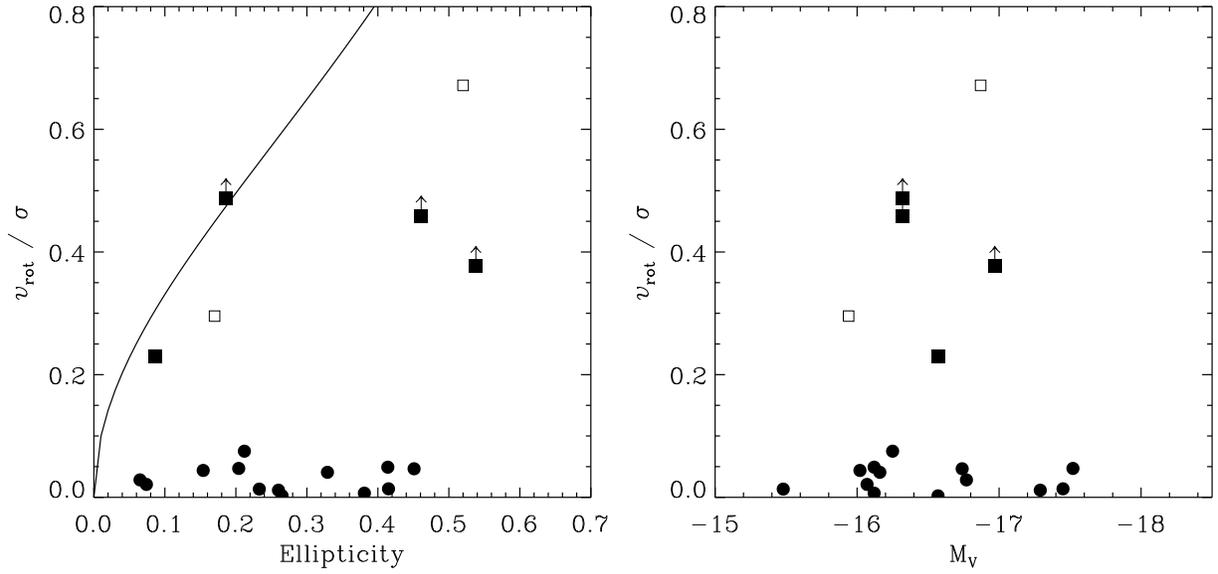}
\vskip 1 cm
\caption{The ratio of the rotation velocity $v_{\rm rot}$ to velocity
dispersion $\sigma$ plotted versus mean isophotal ellipticity ({\it
left panel\/}) and absolute magnitude ({\it right panel\/}).  The
solid line in the left panel is the expected relation for an oblate,
isotropic galaxy flattened by rotation.  Solid symbols indicate Virgo
Cluster dEs from the sample in this paper, while open symbols are two
dEs taken from \citet{sim02}; dEs are plotted as circles if $v_{\rm
rot}/\sigma \le 0.1$ (most of these represent upper limits), and as
squares otherwise.  Lower limits are
indicated for rotating galaxies for which we do not observe a turnover
in the rotation curve due to insufficient radial coverage.   
\label{fig_vsigma}}
\end{figure}

\begin{figure}
\epsscale{0.5}
\plotone{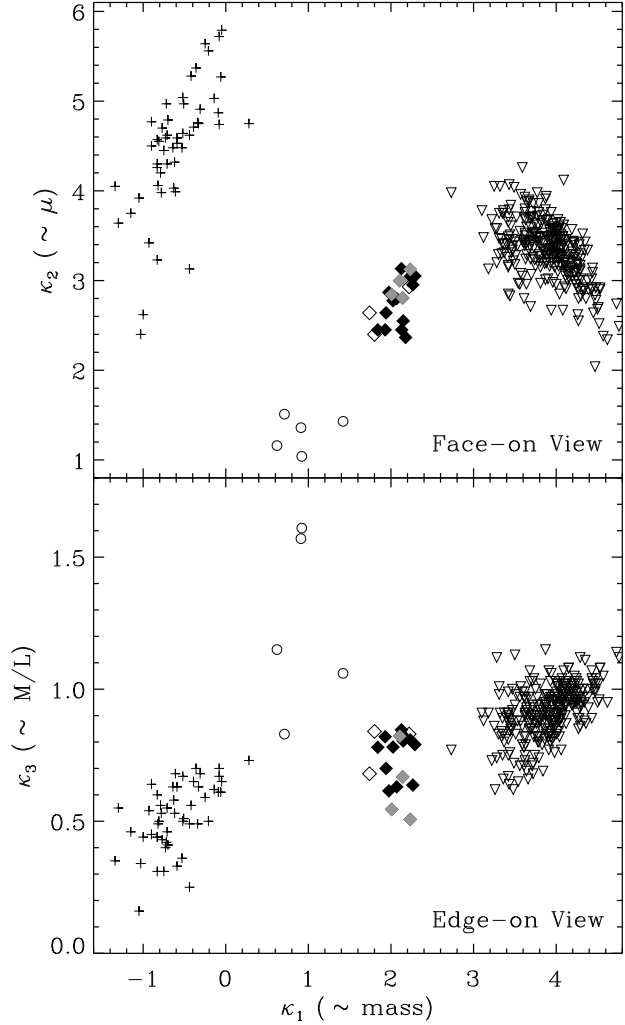}
\caption{Face-on and edge-on projections of the Fundamental Plane for
dynamically hot stellar systems ({\it upper and lower panels,
respectively\/}), where $\kappa_1$, $\kappa_2$, and $\kappa_3$ are
related to galaxy mass, surface brightness, and mass-to-light ratio,
respectively.  Note that the vertical scale is greatly expanded in the lower
panel relative to the upper
panel.  Our sample of dEs is shown as solid diamonds for non-rotating
dEs, and solid grey diamonds for rotating dEs.  Data for other systems
from \citet{bur97} are: classical ellipticals and spiral bulges (open
triangles), Local Group dEs (open diamonds), Local Group dwarf
spheroidals (open circles), and Galactic globular clusters (crosses).
Note the change in vertical scale between the two
panels.\label{fig_fp}}
\end{figure}

\begin{figure}
\epsscale{1.0}
\plotone{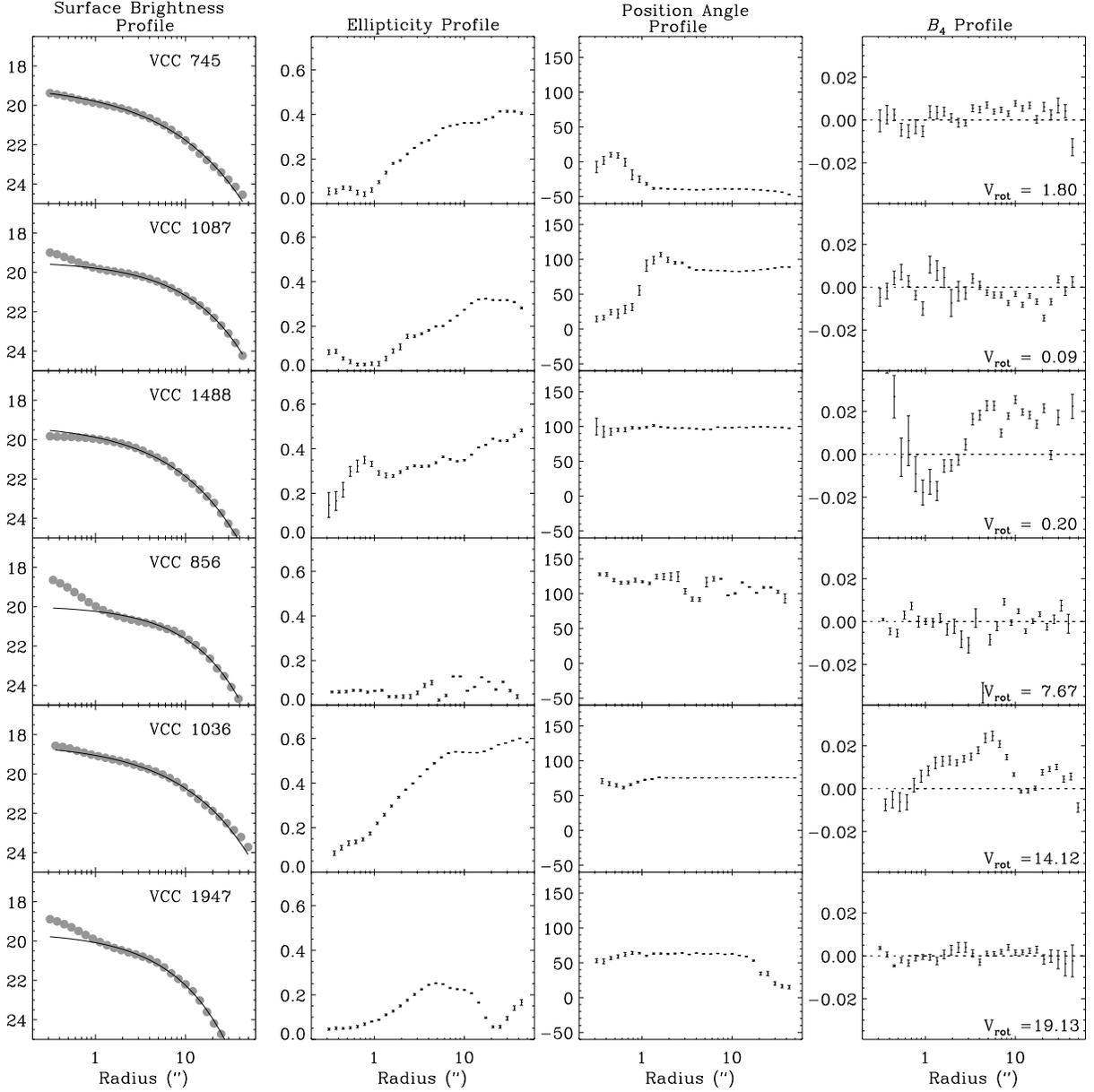}
\caption{For dEs with ESI imaging, $V$-band surface brightness (mag
arcsecond$^{-2}$), ellipticity, position angle (degrees), and $B_4$
deviations from perfect ellipticity are plotted as a function of
radius.  The rotation velocity (\kms) of each galaxy is indicated at
the bottom of the fourth panel in each row; the top three rows are
non-rotating dEs, while the bottom three rows are rotating.  The
solid line in the first column is a S\'{e}rsic profile fitted in the
region $1''-20''$, excluding any nuclear component.
\label{fig_sbesi}} 
\end{figure}

\begin{figure}
\epsscale{0.7}
\plotone{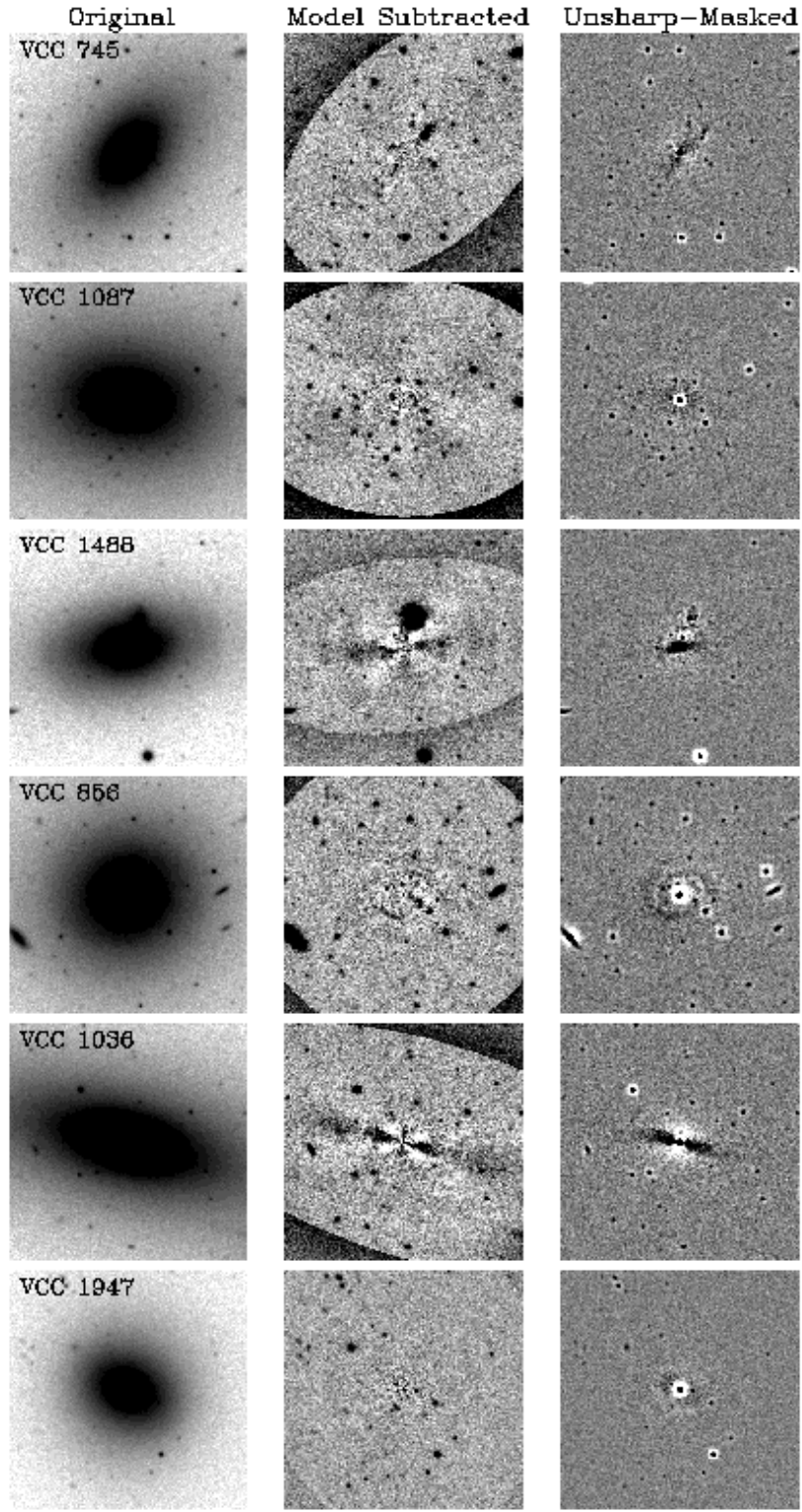}
\caption{ESI imaging data for the same set of dEs
shown in Figure~\ref{fig_sbesi}: ({\it left column\/}) original $V$-band
image; ({\it middle column\/}) residual image after subtracting a
two-dimensional model based on ellipse-fitting of the original image; and
({\it right column\/}) residual image after subtracting a smoothed version
of the original image from itself.  Each panel covers a $1' \times 1'$
region centered on the galaxy.
\label{fig_imesi}} 
\end{figure}

\begin{figure}
\epsscale{0.9}
\plotone{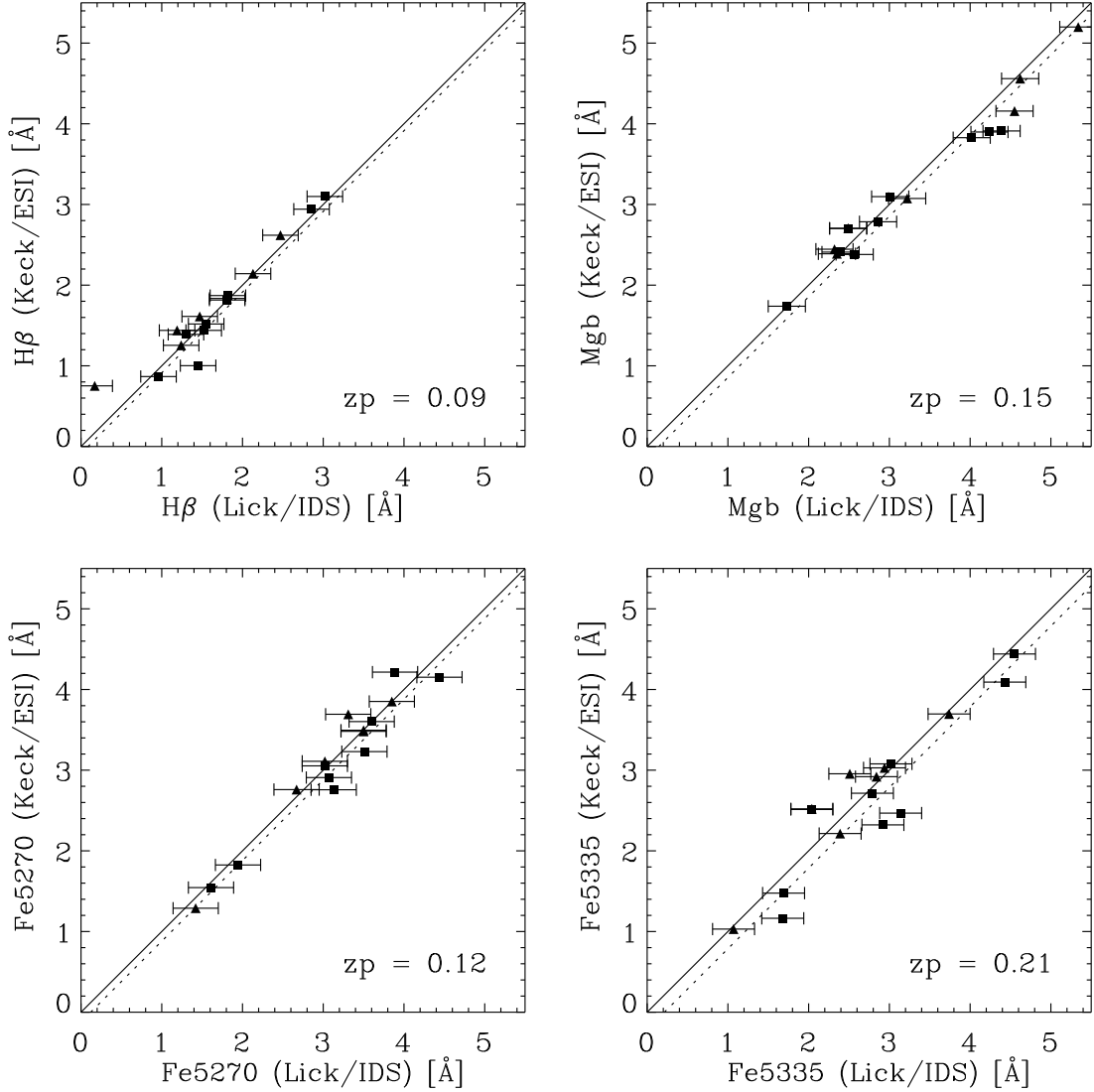}
\caption{Standard star line-strength indices measured with Keck/ESI
compared to the corresponding published Lick/IDS values from
\citet{wor94}.  Square symbols represent data from our March 2001 observing run
\citep{geh02} and triangles represent the observations discussed in
\S\,\ref{sec_spec}.  The solid line has unit slope and zero offset, while
the dashed line has unit slope and the best-fit zero-point correction.
This correction has been applied to our measured indices and is listed
in the lower right corner of each panel. \label{fig_idscal}}
\end{figure}

\begin{figure}
\epsscale{1.0}
\hskip -1 cm
\plotone{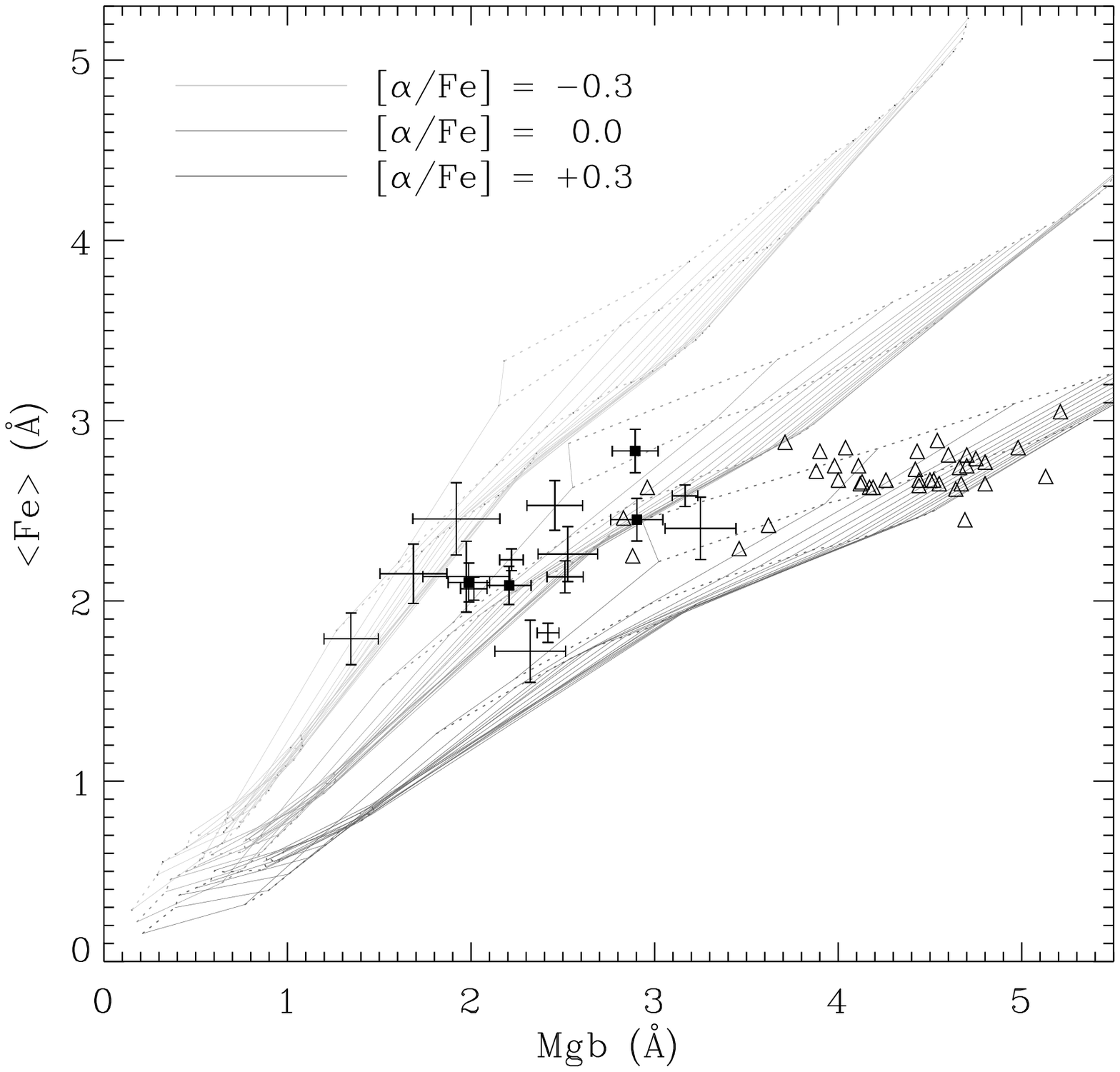}
\caption{Determination of $[\alpha/\rm Fe]$ ratios from a Mgb versus
$\langle\rm Fe\rangle$ diagram.  The dEs in our sample are plotted as
$1\sigma$ error crosses; dEs with significant rotation velocities are
shown with solid squares.  The classical elliptical galaxy sample of
\citet{tra00} is plotted as open triangles.  Model predictions by
\citet{tho02} are plotted for the abundance ratios $[\alpha/\rm Fe] =
-0.3$, 0.0, and~+0.3 (light to dark grey lines), age = $1-15$~Gyr in
increments of 1~Gyr, and $[\rm Fe/H] = -2.25$, $-1.35$, $-0.33$, 0.0,
$+0.35$, and $+0.67$\,dex.  Rotating and non-rotating dEs cannot be
distinguished in this plot.  The majority of these dEs are consistent
with solar abundance ratios, in contrast with the majority of
classical elliptical galaxies which have enhanced $[\alpha/\rm Fe]$
abundance ratios.
\label{fig_alpha}}
\end{figure}

\begin{figure}
\epsscale{1.0}
\plotone{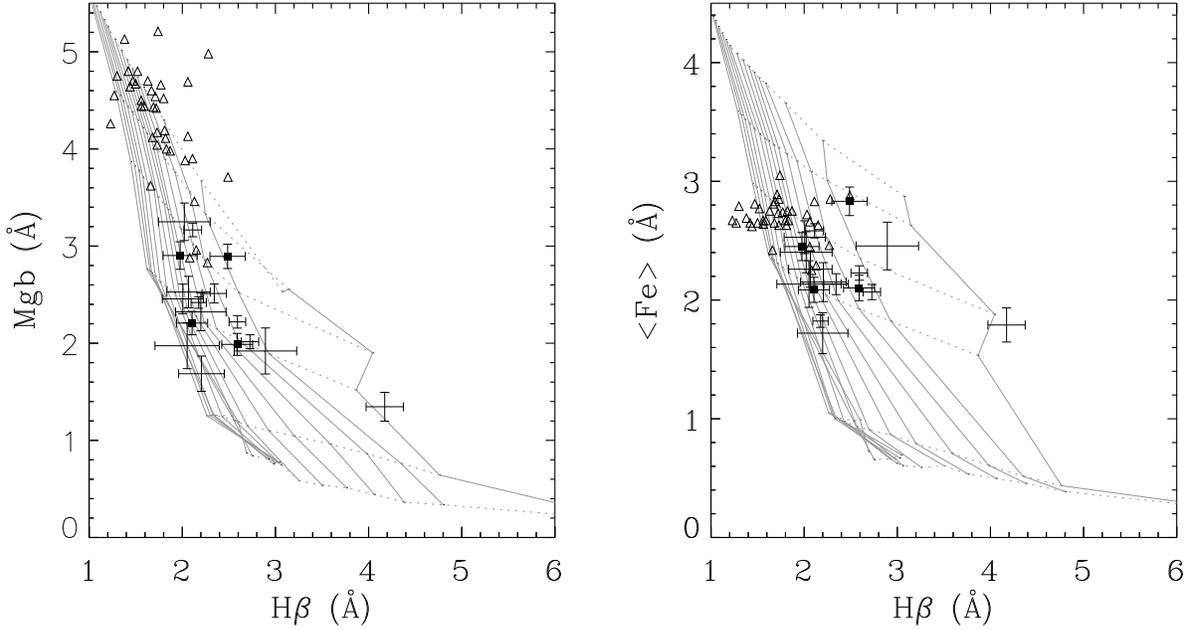}
\vskip 1 cm
\caption{Line-strengths Mgb and $\langle\rm Fe\rangle$ plotted against
H$\beta$ ({\it left and right panels, respectively\/}).  The dEs in
our sample are plotted as $1\sigma$ error crosses.  Dwarf ellipticals
with significant rotation velocities are indicated with solid squares;
as in previous figures, these galaxies cannot be distinguished from
the non-rotating dEs.  The solar abundance ratio models of
\citet{tho02} are plotted for age between $1-15$~Gyr in 1~Gyr
intervals (solid grey lines) and $[\rm Fe/H] = -2.25$, $-1.35$,
$-0.33$, 0.0, $+0.35$, and $+0.67$\,dex (dotted lines).  In both
panels, lines of constant age are steeper than those of constant
metallicity.  The average absolute age and metallicity for our dEs are
5~Gyr and $-0.3$~dex, respectively, determined as described in
\S\,\ref{sec_ids}.  The galaxy with the largest H$\beta$ value is
VCC~1488; excluding this dE in determining the mean age
and metallicity does not significantly affect the values given above.
The classical elliptical galaxy sample of
\citet{tra00} is plotted as open triangles. The average dE has
stronger H$\beta$ and weaker Mgb and $\langle \rm Fe \rangle$ than the
typical elliptical galaxy.\label{fig_ids}}
\end{figure}

\begin{figure}
\epsscale{0.9}
\vskip -1 cm
\plotone{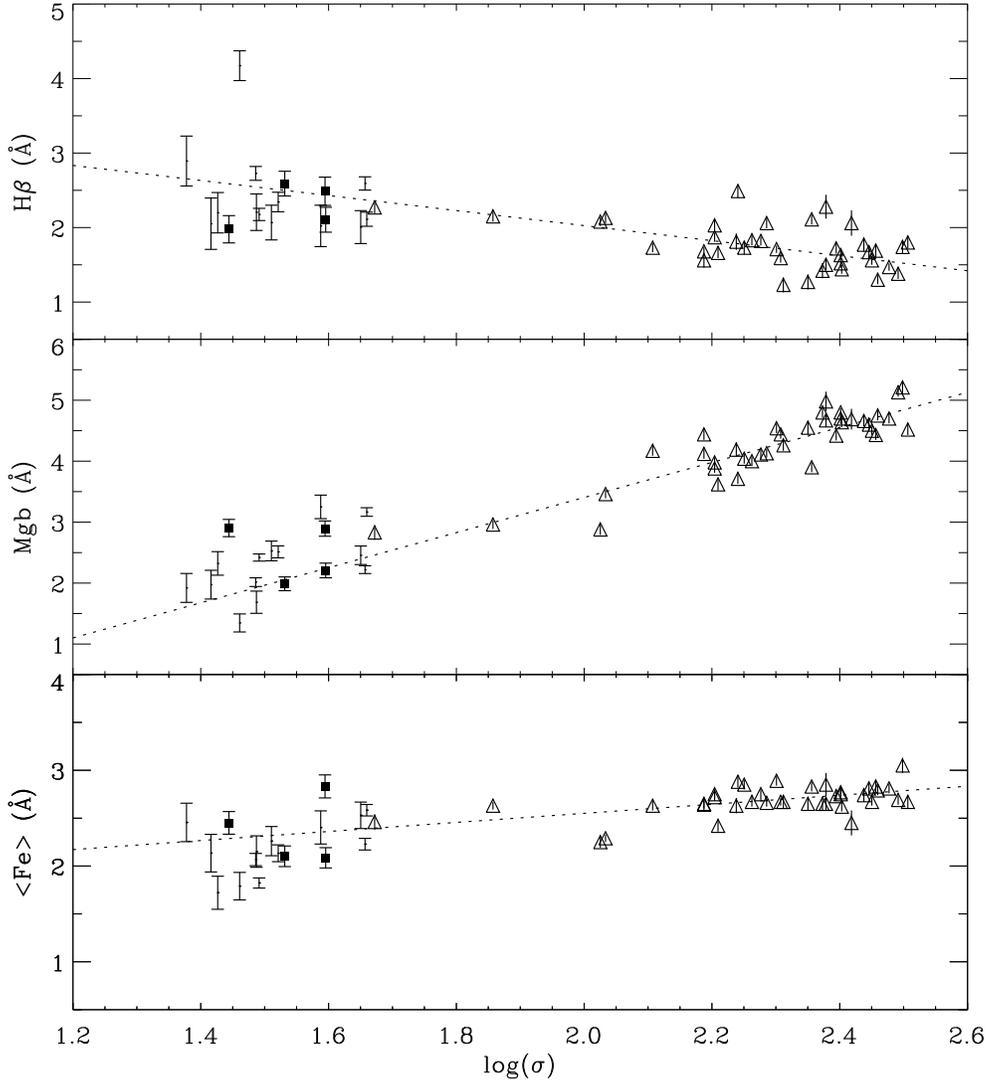}
\caption{Line-strength indices plotted as a function of the logarithm
of the average velocity dispersion $\sigma$.  Symbols in each panel are the
same as in Figures~\ref{fig_alpha}~and~\ref{fig_ids}.  Dotted lines are
fits to the classical ellipticals (open triangles) of \citet{tra00}.
The extrapolation of these fits are consistent with our measured dE
line-strengths.\label{fig_mgsig}}
\end{figure}

\begin{figure}
\epsscale{1.05}
\plotone{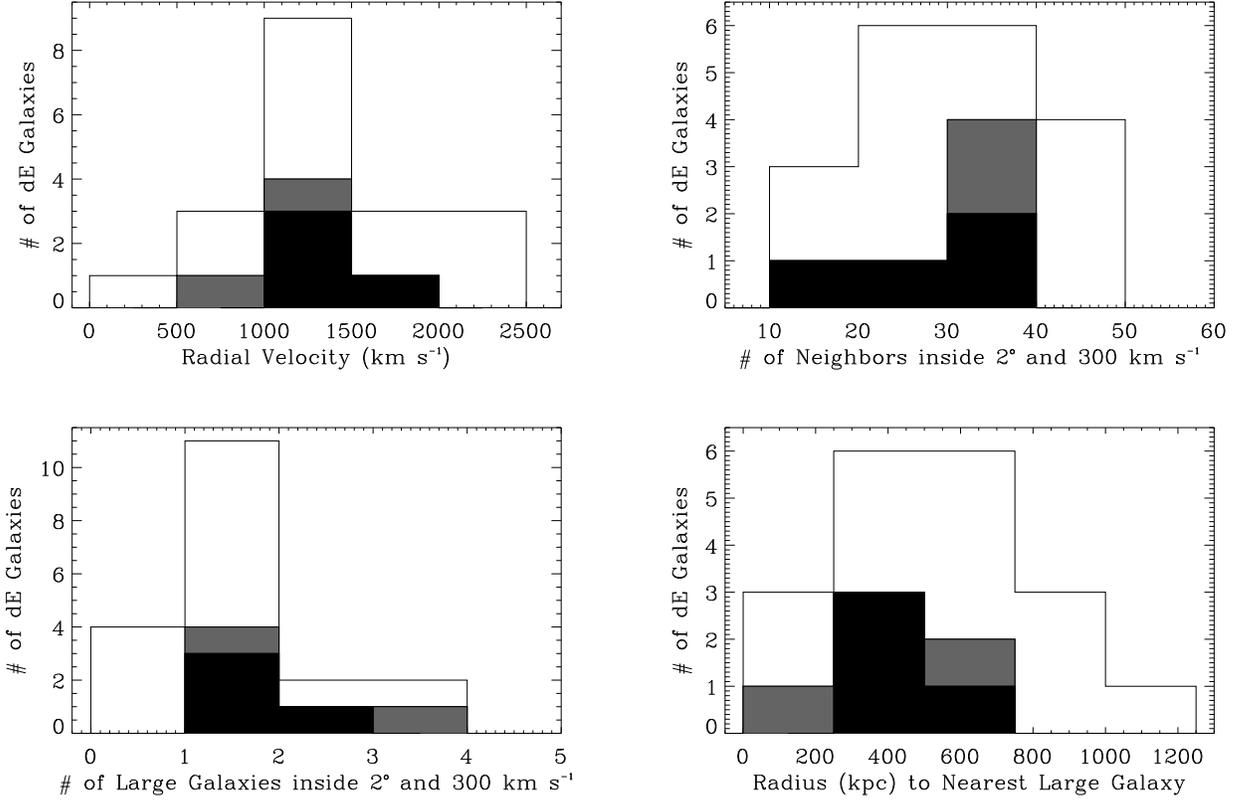}
\vskip 0.5 cm
\caption{The local environment of Virgo dEs.  In each panel, the black
shaded portion of the histogram represents dEs with significant
rotation velocity, the grey shaded region includes rotating dEs from
the literature, the clear region includes all dEs with measured
internal kinematics in the Virgo Cluster.  ({\it Upper left panel\/})
Distribution of radial velocities, ({\it upper right panel\/}) number
of neighboring galaxies inside a radius of $2^{\circ}$ (0.5\,Mpc) and
within 300\kms, ({\it lower left panel\/}) number of galaxies
brighter than $M_V = -20$ in the same region, and ({\it lower right
panel\/}) radius in kpc to the nearest of these bright galaxies.
\label{fig_environ}}
\end{figure}

In the above sections we have demonstrated that dEs with and without
significant rotation velocities cannot be distinguished based on their
internal properties.  We next examine whether or not internal rotation
is correlated with a dE's local environment within the Virgo Cluster.
As shown in the left panel of Figure~\ref{fig_vcc}, rotating and
non-rotating dEs are found at varying radii from the center of Virgo.
In velocity space (Figure~\ref{fig_vcc}, right panels), the
non-rotating dEs span the full range of radial velocities found in the
Virgo, whereas the rotating dEs appear clustered near the average
Virgo radial velocity of $\sim 1000$\kms.  This is more clearly
evident in the first panel of Figure~\ref{fig_environ}.  Based on the
K-S test discussed in \S\,\ref{sec_fp}, the radial velocity
distributions of the rotating versus non-rotating dEs are not
significantly different ($P_{\rm KS} = 0.3$), although including two
additional dEs from the literature \citep{sim02} makes this difference
marginally significant ($P_{\rm KS}=0.06$) in the sense that rotating
dEs have a narrower range in radial velocities.  \citet{con01} have
argued that Virgo Cluster dEs as a population span a much larger range
of radial velocities as compared to more luminous classical
ellipticals.  They interpret this as evidence that dEs are a more
recently accreted population.  If larger dE samples confirm that
rotating dEs span a narrower range of radial velocities compared to
non-rotating dEs, this would suggest that their history in the Virgo
Cluster is more closely associated with classical elliptical galaxies.

In the remaining panels of Figure~\ref{fig_environ}, we further
quantify the local environment of our dE sample by examining
neighboring galaxies within $2^{\circ}$ (0.5\,Mpc) and 300\,\kms\ of
each dE.  These scales were chosen to probe local enhancements within
the Virgo Cluster on the scale of small galaxy groups.  As a
population, Virgo dEs are not preferentially found around larger
parent galaxies \citep{fer91} in contrast to dEs in the field; local
overdensities or proximity to large galaxies would be evidence of
recent accretion into the Virgo Cluster from the surrounding
field/groups.  We compare for the rotating and non-rotating dEs: the
number of neighboring galaxies contained within the physical distance
and velocity difference stated above, the number of luminous
companions ($M_V < -20$) in this same region, and the physical
distance to the nearest of these bright galaxies.  In all cases, the
rotating and non-rotating dEs are similarly distributed based on both
inspection of Figure~\ref{fig_environ} and results of the K-S test.
In all three cases the K-S probability is greater than $P_{\rm KS} =
0.25$, meaning that the distribution of the two kinematic types in
these panels cannot be distinguished.  We conclude that internal dE
rotation properties are not correlated with proximity to bright
galaxies or the local environment within the galaxy cluster.

\section{Discussion and Conclusions}\label{sec_discuss}

We have established that the internal dynamics of dE galaxies span a
wide range in rotational support as determined from major axis
spectroscopy of 17 dEs in the Virgo Cluster.  In our present dE
sample, the degree of rotational support does not appear to span a continuous
distribution: we observe either dEs with no measurable major axis
rotation, or dEs with rotation velocities approaching that expected
for a rotationally-flattened object.  We therefore have separated our
sample into rotating and non-rotating dEs (nominally defined as
galaxies with $v_{\rm rot}/\sigma > 0.1$ and $v_{\rm rot}/\sigma \le
0.1$, respectively) and investigated whether or not other dE
properties are correlated with rotation
velocity.  We show that rotation velocity is not correlated with a dE's
position in the Fundamental Plane, the presence or absence of
underlying disks or substructure, absorption-line-strength indices, or local
environment.

Currently favored formation models do not naturally predict a
dichotomy in the internal dynamics of dE galaxies.  One possible
explanation for the varying amount of rotational support is
different mechanisms for the formation of rotating and non-rotating
dEs.  Although our observations do not rule out multiple formation
mechanisms, they do place very tight constraints on variations in observable
properties between
the resulting dE populations.  In particular, significantly different
origins for rotating and non-rotating dEs would need to result in
stellar populations that are indistinguishable between the two dE types yet
very different from those of more luminous classical ellipticals.  It is
equally challenging however to produce the range in rotational
properties via a single physical process.

Our conclusion that the internal properties and local environments of
rotating versus non-rotating cannot be distinguished is limited by the
size of our dE sample; increasing the number of dEs, particularly rotating
dEs, with spatially-resolved measurements of
internal kinematics is an important step towards further
understanding the formation of these galaxies.  It is equally
important to refine the predictions and better understand the limitations of
currently proposed dE formation models.  For example, does the
harassment scenario in which dEs are formed from larger disk
galaxies via gravitational interactions produce the observed range of 
rotational support, and, in particular, can it account for the large
fraction of dEs without significant rotation?  To what extent do these
interactions disrupt the disk of a progenitor galaxy and at what level
is substructure predicted in dEs?  Similar questions should be asked
of any proposed dE formation model.  Because dEs are so numerous in
nearby galaxy clusters, constraining the formation and evolution of
these objects is an important step towards assembling a global picture
of galaxy evolution in these regions.

\acknowledgments

We thank Ricardo Schiavon and Sandy Faber for fruitful discussions
regarding this work.  M.G.\ acknowledges a Presidential Dissertation
Year Fellowship from the University of California at Santa Cruz and support
from the STScI Director's Discretionary Research Fund.
P.G.\ would like to thank Joe Miller for his support through a UCO/Lick
Observatory Director's grant, and acknowledges partial support from
NASA/STScI grants GO-08192.03A and GO-08601.04A.




\clearpage

\begin{deluxetable}{lcclcccccc}
\tabletypesize{\scriptsize} 
\tablecaption{Basic Properties}
\tablewidth{0pt} 
\tablehead{ 
\colhead{Galaxy} & 
\colhead{$\alpha$(J2000)} & 
\colhead{$\delta$ (J2000)} & 
\colhead{Type} &
\colhead{$m_V$} & 
\colhead{$M_{V,0}$}& 
\colhead{$v_{\rm sys}$}&
\colhead{$v_{\rm rot}$}& 
\colhead{$\sigma$}& 
\colhead{Ref}\\
\colhead{}& 
\colhead{(h$\,$:$\,$m$\,$:$\,$s)} &
\colhead{($^\circ\,$:$\,'\,$:$\,''$)} &
 \colhead{} & \colhead{} &
\colhead{}& 
\colhead{(\kms)}& 
\colhead{(\kms)}& 
\colhead{(\kms)}&
\colhead{} 
} 
\startdata 
\multicolumn{4}{l} {\bf Non-Rotating Dwarf Elliptical Galaxies}\\ 
VCC 452           & 12:21:04.7 &  11:45:18 & dE4N  & 15.34 & $-16.02$ & 1380 &  $1.0\pm0.5$ & $23.8\pm1.0$ & 1\\
VCC 745/NGC~4366  & 12:24:47.0 &  07:21:10 & dE6N  & 14.74 & $-16.16$ & 1234 &  $1.8\pm0.4$ & $44.7\pm0.8$ & 2\\
VCC 917/IC~3344   & 12:26:32.4 &  13:34:43 & dE6   & 13.93 & $-17.45$ & 1186 &  $0.4\pm0.4$ & $30.6\pm0.4$ & 1\\
VCC 940/IC~3349   & 12:26:47.1 &  12:27:15 & dE1N  & 14.83 & $-16.07$ & 1563 &  $0.7\pm0.7$ & $32.4\pm1.7$ & 1\\
VCC 1073/IC~794   & 12:28:08.6 &  12:05:36 & dE3N  & 13.84 & $-17.52$ & 1862 &  $2.1\pm0.5$ & $45.6\pm0.3$ & 2,3\\
VCC 1087/IC~3381  & 12:28:15.1 &  11:47:23 & dE3N  & 14.33 & $-16.57$ &  650 &  $0.5\pm0.4$ & $38.7\pm1.1$ & 1,4\\
VCC 1254          & 12:30:05.3 &  08:04:29 & dE0N  & 14.58 & $-16.77$ & 1220 &  $0.9\pm0.9$ & $31.0\pm0.9$ & 2\\
VCC 1261/NGC~4482 & 12:30:10.4 &  10:46:46 & dE5N  & 13.61 & $-17.29$ & 1845 &  $0.5\pm0.4$ & $45.4\pm0.5$ & 1,4\\
VCC 1308/IC~3437  & 12:30:45.8 &  11:20:34 & dE6N  & 15.42 & $-15.48$ & 1779 &  $0.4\pm0.4$ & $33.2\pm0.6$ & 1\\
VCC 1386/IC~3457  & 12:31:51.3 &  12:39:21 & E3:N  & 14.65 & $-16.25$ & 1150 &  $2.3\pm0.5$ & $30.7\pm2.3$ & 1\\
VCC 1488/IC~3487  & 12:33:13.4 &  09:23:50 & E6:\tablenotemark{a}   & 14.78 & $-16.12$ & 1157 &  $0.5\pm0.3$ & $28.9\pm0.8$ & 1\\ 
VCC 1577/IC~3519  & 12:34:38.4 &  15:36:10 & dE4   & 15.24 & $-16.12$ &  361 &  $1.3\pm0.7$ & $26.0\pm0.8$ & 2\\
VCC 1876/IC~3658  & 12:41:20.4 &  14:42:02 & dE5N  & 14.64 & $-16.74$ &   95 &  $1.2\pm1.4$ & $26.7\pm1.3$ & 2\\ \\
\multicolumn{4}{l} {\bf Rotating Dwarf Elliptical Galaxies}\\ 
VCC 543/UGC~7436  & 12:22:19.5 &  14:45:39 & dE5   & 14.58    & $-16.32$ &  923 & $>12.9\pm1.9$ & $27.8\pm0.4$ & 1,3,4\\
VCC 856/IC~3328   & 12:25:57.8 &  10:03:13 & dE1N  & 14.33    & $-16.57$ & 1014 &  ~~~~~$7.7\pm1.4$ & $34.0\pm1.2$ & 1,4\\
VCC 1036/NGC~4436 & 12:27:41.6 &  12:18:59 & dE6/dS0N & 13.93 & $-16.97$ & 1324 & $>14.1\pm1.7$ & $39.3\pm0.7$ & 1,4\\
VCC 1947          & 12:42:56.3 &  03:40:35 & dE2N  & 14.58    & $-16.32$ & 1083 & $>19.1\pm2.8$ & $39.4\pm0.8$ & 1\\
\enddata 
\tablenotetext{a}{Although \citet{bin85} classify VCC~1488 as an ``E6:'',
its nearly exponential surface brightness profile and position in the Fundamental Plane
suggest the correct classification is dE.}
\tablecomments{Galaxy classifications are taken from
\citet{bin85}; object types including ``:'' indicate uncertainty in
the classification.  Apparent and absolute magnitudes are determined
inside an $r<40''$ aperture.  The absolute magnitude, $M_{V,0}$,
assumes a Virgo Cluster distance modulus of $(m - M)_0 = 30.92$ and is
corrected for foreground dust extinction according to \citet{sch98}. The
heliocentric systemic velocity $v_{\rm sys}$ is determined from the
mean of each velocity profile.  The rotation speed $v_{\rm rot}$ and
associated error bar are determined as described in \S\,\ref{sec_kin}.
The average line-of-sight velocity dispersion $\sigma$ is determined
for $r>1''$ to avoid nuclear contamination if present.  The final
column refers to published kinematic observations as follows: 1 = this
paper, 2 = \citet{geh02}, 3 = \citet{ped02}, and 4 = \citet{sim02}.}
\end{deluxetable}

\begin{deluxetable}{lcrccccc}
\tabletypesize{\scriptsize}
\tablecaption{Parameters for New Spectroscopic Observations}
\tablewidth{0pt}
\tablehead{
\colhead{Galaxy} &
\colhead{Exposure Time} &
\colhead{PA$_{\rm slit}$}&
\colhead{Seeing FWHM} \\
\colhead{}&
\colhead{(s)}&
\colhead{($^{\circ}$)} & 
\colhead{($''$)} 
} 
\startdata
VCC 543        & 1800 & $-54$~~~ & 0.9 \\
VCC 745        & 1600 &   45~~~  & 0.8 \\
VCC 856        & 1500 &   80~~~  & 0.7 \\
VCC 940        & 3600 &   0~~~   & 0.7 \\
VCC 1036       & ~900 & $-68$~~~ & 0.8 \\
VCC 1087       & 1500 & $-77$~~~ & 0.8 \\
VCC 1261       & 2700 & $-50$~~~ & 0.8 \\
VCC 1308       & 3600 &  60~~~   & 0.7 \\
VCC 1386       & 3600 & $-40$~~~ & 0.9 \\
VCC 1488       & 1500 & 80~~~    & 0.7 \\
VCC 1947       & ~900 & $-65$~~~ & 0.8 \\
\enddata
\tablecomments{Exposure time, slit position angle and seeing FWHM for
the eleven dEs whose Keck/ESI kinematics are presented in this paper.
The slit was placed on the major axis of each galaxy.  Seeing FWHMs
were determined from guider camera images taken simultaneously with
the observations.}
\end{deluxetable}

\begin{deluxetable}{lclccccccc}
\tabletypesize{\scriptsize}
\tablecaption{Photometric Properties} 
\tablewidth{0pt}
\tablehead{
\colhead{Galaxy} &
\colhead{Imaging}&
\colhead{$\epsilon$} &
\colhead{$n_{\rm Sersic}$} &
\colhead{$r_{\rm eff}$} &
\colhead{$\mu_{V,\rm eff}$}&
\colhead{$\kappa_1$}&
\colhead{$\kappa_2$}&
\colhead{$\kappa_3$}&
\colhead{$B_4 \times 100$}\\
\colhead{}&
\colhead{}&
\colhead{}&
\colhead{}&
\colhead{[$''$ (kpc)]}&
\colhead{(mag arcsec$^{-2}$)}&
\colhead{}&
\colhead{}&
\colhead{}&
\colhead{}
}
\startdata
\multicolumn{4}{l} {\bf Non-Rotating Dwarf Elliptical Galaxies}\\
VCC           452 & WFPC2 &  0.15 &  1.6 &  ~9.6 (0.71) &  22.3 &   1.84 &   2.45 &   0.78 & ~~$0.13\pm 0.28$\\
VCC  745/NGC~4366 &  ESI &  0.33 &  1.5 &  10.0 (0.74) &  21.1 &   2.24 &   3.05 &   0.81 & ~~$0.35\pm 0.03$\\
VCC  917/IC~3344  & WFPC2 &  0.42 &  2.9 &  12.2 (0.90) &  21.4 &   2.07 &   2.80 &   0.63 & ~~$0.34\pm 0.18$\\
VCC  940/IC~3349  & WFPC2 &  0.07 &  1.3 &  13.9 (1.03) &  22.2 &   2.15 &   2.55 &   0.80 & ~~$0.34\pm 0.22$\\
VCC 1073/IC~794   & WFPC2 &  0.20 &  1.9 &  11.1 (0.82) &  21.1 &   2.29 &   3.05 &   0.79 & $-0.03\pm 0.16$\\
VCC 1087/IC~3381  &  ESI &  0.26 &  1.4 &  14.4 (1.07) &  21.1 &   2.26 &   2.95 &   0.64 & $-0.41\pm 0.03$\\
VCC 1254          & WFPC2 &  0.07 &  2.9 &  14.4 (1.07) &  22.4 &   2.13 &   2.45 &   0.83 & $-0.60\pm 0.13$\\
VCC 1261/NGC~4482 & WFPC2 &  0.26 &  1.9 &  ~7.3 (0.54) &  21.0 &   2.13 &   3.13 &   0.85 & ~~$0.54\pm 0.11$\\
VCC 1308/IC~3437  & WFPC2 &  0.23 &  1.3 &  ~9.5 (0.70) &  21.7 &   2.02 &   2.77 &   0.78 & ~~$0.06\pm 0.20$\\
VCC 1386/IC~3457  & WFPC2 &  0.21 &  1.3 &  17.1 (1.27) &  22.5 &   2.18 &   2.37 &   0.81 & $-0.46\pm 0.25$\\
VCC 1488/IC~3487  &  ESI &  0.38 &  1.6 &  10.0 (0.74) &  21.2 &   1.97 &   2.87 &   0.62 & ~~$1.40\pm 0.05$\\ 
VCC 1577/IC~3519  & WFPC2 &  0.41 &  1.1 &  10.5 (0.78) &  22.4 &   1.93 &   2.45 &   0.82 & ~~$0.52\pm 0.49$\\
VCC 1876/IC~3658  & WFPC2 &  0.45 &  0.8 &  10.5 (0.78) &  21.8 &   1.94 &   2.64 &   0.70 & ~~$0.61\pm 0.30$\\ \\
\multicolumn{5}{l} {\bf Rotating Dwarf Elliptical Galaxies}\\
VCC  543/UGC~7436 & WFPC2 &  0.46 &  1.4 &  11.9 (0.88) &  21.2 &   2.01 &   2.84 &   0.54 & $-0.36\pm 0.26$\\
VCC  856/IC~3328  &  ESI &  0.09 &  1.6 &  13.0 (0.96) &  21.4 &   2.14 &   2.81 &   0.67 & ~~$0.04\pm 0.03$\\
VCC 1036/NGC~4436 &  ESI &  0.54 &  1.5 &  13.8 (1.02) &  20.6 &   2.23 &   3.12 &   0.51 & ~~$0.65\pm 0.03$\\
VCC 1947          &  ESI &  0.19 &  1.3 &  ~8.3 (0.62) &  21.3 &   2.11 &   3.00 &   0.82 & $-0.05\pm 0.02$\\
\enddata 
\tablecomments{The second column refers to the source of our imaging
(HST/WFPC2 or Keck/ESI).  The ellipticity $\epsilon$ is the average
measured between $1''<r<20''$.  The S\'{e}rsic index $n_{\rm Sersic}$,
effective (half-light) radius $r_{\rm eff}$, and effective surface
brightness $\mu_{V,\rm eff}$ are determined by fitting a S\'{e}rsic
model to the galaxy surface brightness profile outside $r>1''$ where
the radius is measured along the major axis.  The Fundamental Plane
parameters $\kappa_1$, $\kappa_2$, and $\kappa_3$ were determined from
quantities in this table and Table~1 according to the definitions of
\citet{ben92}.  The $\kappa$-values are defined in the $B$-band, and
we assume $B-V$ = 0.8 in calculating these parameters.  The last
column lists the disky/boxy parameter $B_4$ (expressed as a
percentage) discussed in \S\,\ref{sec_morph} determined over the
radial range $1''<r<20''$.}

\end{deluxetable}

\begin{deluxetable}{lccccc}
\tabletypesize{\scriptsize}
\tablecaption{Line-Strength Indices} 
\tablewidth{0pt}
\tablehead{
\colhead{Galaxy} &
\colhead{H$\beta$}&
\colhead{Mgb}&
\colhead{Fe5270}&
\colhead{Fe5335}&
\colhead{$\langle \rm Fe\rangle$} \\
\colhead{}&
\colhead{($\mbox{\AA}$)}&
\colhead{($\mbox{\AA}$)}&
\colhead{($\mbox{\AA}$)}&
\colhead{($\mbox{\AA}$)}&
\colhead{($\mbox{\AA}$)}
}
\startdata
\multicolumn{4}{l} {\bf Non-Rotating Dwarf Elliptical Galaxies}\\
VCC  452  & $2.89 \pm 0.34$ & $1.92 \pm 0.24$ & $2.65 \pm 0.27$ & $2.26 \pm 0.29$ & $2.46 \pm 0.20$ \\ 
VCC  745  & $2.01 \pm 0.22$ & $2.46 \pm 0.15$ & $2.76 \pm 0.19$ & $2.30 \pm 0.20$ & $2.53 \pm 0.14$ \\
VCC  917  & $2.73 \pm 0.09$ & $2.02 \pm 0.07$ & $2.38 \pm 0.09$ & $1.76 \pm 0.09$ & $2.07 \pm 0.06$ \\
VCC  940  & $2.07 \pm 0.23$ & $2.53 \pm 0.16$ & $2.46 \pm 0.21$ & $2.06 \pm 0.22$ & $2.26 \pm 0.15$ \\
VCC 1073  & $2.11 \pm 0.10$ & $3.17 \pm 0.07$ & $2.80 \pm 0.09$ & $2.36 \pm 0.08$ & $2.58 \pm 0.06$ \\
VCC 1087  & $2.02 \pm 0.28$ & $3.25 \pm 0.19$ & $2.62 \pm 0.23$ & $2.18 \pm 0.26$ & $2.40 \pm 0.17$ \\
VCC 1254  & $2.18 \pm 0.08$ & $2.42 \pm 0.06$ & $2.09 \pm 0.07$ & $1.55 \pm 0.08$ & $1.82 \pm 0.05$ \\
VCC 1261  & $2.59 \pm 0.09$ & $2.22 \pm 0.06$ & $2.33 \pm 0.09$ & $2.12 \pm 0.09$ & $2.23 \pm 0.06$ \\
VCC 1308  & $2.34 \pm 0.13$ & $2.51 \pm 0.10$ & $2.41 \pm 0.12$ & $1.86 \pm 0.13$ & $2.13 \pm 0.09$ \\
VCC 1386  & $2.21 \pm 0.24$ & $1.69 \pm 0.18$ & $2.35 \pm 0.23$ & $1.95 \pm 0.23$ & $2.15 \pm 0.16$ \\
VCC 1488  & $4.17 \pm 0.20$ & $1.35 \pm 0.15$ & $1.78 \pm 0.19$ & $1.80 \pm 0.22$ & $1.79 \pm 0.14$ \\
VCC 1577  & $2.05 \pm 0.34$ & $1.97 \pm 0.24$ & $2.58 \pm 0.26$ & $1.69 \pm 0.29$ & $2.13 \pm 0.20$ \\
VCC 1876  & $2.20 \pm 0.27$ & $2.32 \pm 0.19$ & $2.06 \pm 0.23$ & $1.38 \pm 0.26$ & $1.72 \pm 0.17$ \\ \\
\multicolumn{4}{l} {\bf Rotating Dwarf Elliptical Galaxies}\\ 
VCC  543  & $1.98 \pm 0.18$ & $2.90 \pm 0.14$ & $2.56 \pm 0.16$ & $2.34 \pm 0.17$ & $2.45 \pm 0.12$ \\
VCC  856  & $2.59 \pm 0.17$ & $1.99 \pm 0.11$ & $2.23 \pm 0.15$ & $1.98 \pm 0.16$ & $2.10 \pm 0.11$ \\
VCC 1036  & $2.49 \pm 0.19$ & $2.89 \pm 0.12$ & $2.85 \pm 0.17$ & $2.82 \pm 0.17$ & $2.83 \pm 0.12$ \\
VCC 1947  & $2.10 \pm 0.17$ & $2.21 \pm 0.12$ & $2.58 \pm 0.15$ & $1.59 \pm 0.15$ & $2.09 \pm 0.11$ \\
\enddata

\tablecomments{Line-strength indices are determined according to the
definitions of \citet{wor94}.  Error bars are
computed via Monte-Carlo simulations as described in
\S\,\ref{sec_ids}.  The combined iron index is defined as $\langle \rm
Fe\rangle \equiv (\rm Fe5270 + \rm Fe5335) / 2$.}
\end{deluxetable}

\end{document}